\documentclass[10pt,twocolumn,twoside]{IEEEtran}
\usepackage{lineno,hyperref}
\usepackage{setspace}
\usepackage{amsmath}
\usepackage{graphicx}
\usepackage{algorithm}
\usepackage{geometry}
\usepackage{color}
\usepackage{amsmath}
\usepackage{amssymb}
\usepackage{amsfonts}
\usepackage{caption}
\usepackage{subcaption}
\usepackage{algpseudocode}
\usepackage{amsmath}
\usepackage{cases}
\usepackage{multirow}
\usepackage{etoolbox}
\usepackage{graphicx}
 \usepackage{epstopdf}
 \hypersetup{colorlinks=true,linkcolor=black}

\usepackage{stfloats}
 \modulolinenumbers[5]

\newtheorem{lemma}{Lemma}
\BeforeBeginEnvironment{lemma}{\noindent}
\newtheorem{theorem}{Theorem}

\usepackage{epsfig}
\graphicspath{{./plots/}, {./figures/}}

\newcommand{\LOneNormSQ}[1]{{|#1|^2}}
\newcommand{\LTwoNormSQ}[1]{{||#1||_2^2}}
\newcommand{\BigLTwoNorm}[1]{{\Bigl\|#1\Bigr\|_2}}

\newcommand{\VLTwoNormSQ}[1]{\left\lVert#1\right\rVert_2^2}

\singlespace
\IEEEoverridecommandlockouts
\makeatletter
  \def\vhrulefill#1{\leavevmode\leaders\hrule\@height#1\hfill \kern\z@}
\makeatother
 
\begin{document}
\title{Energy Efficient Beamforming Design for MISO Non-Orthogonal Multiple Access Systems}
\author{Haitham Al-Obiedollah,~\IEEEmembership{ Student Member, IEEE},  Kanapathippillai Cumanan, ~\IEEEmembership{Member, IEEE}, Jeyarajan Thiyagalingam, ~\IEEEmembership{Member, IEEE}, Alister G. Burr, ~\IEEEmembership{ Senior Member, IEEE},    Zhiguo Ding, ~\IEEEmembership{Senior Member, IEEE}, and Octavia A. Dobre, ~\IEEEmembership{Senior Member, IEEE}  
 \vspace{-0.3in}
\thanks{H. Alobiedollah, K. Cumanan, and A. G. Burr are with the Department of Electronic Engineering, University of York, York, YO10 5DD, UK. (Email: \{hma534, kanapathippillai.cumanan, alister.burr\}@york.ac.uk.}
\thanks{J. Thiyagalingam is with the Scientific Computing Department
   of Rutherford Appleton Laboratory, Science and Technology
   Facilities Council, Harwell Campus, Oxford, UK. (Email:
   t.jeyan@stfc.ac.uk).}
\thanks{Z. Ding is with the  School of Electrical and Electronic Engineering, The University of Manchester,  
 Manchester, UK.(Email: zhiguo.ding@manchester.ac.uk).}
\thanks {O. A. Dobre is with the Department of Electrical and Computer
Engineering, Memorial University, St. John’s, NL A1B 3X5, Canada (email:
odobre@mun.ca).}
\thanks { The work of H. Alobiedollah was supported by the Hashemite University, Zarqa, Jordan. The work of K. Cumanan, A. G. Burr, and  Z. Ding was supported by H2020-MSCA-RISE-2015 under grant number 690750. The work of Z. Ding was also supported by the UK EPSRC under grant number EP/N005597/2. The work of O. A. Dobre was supported by Natural Sciences and Engineering Research Council of Canada (NSERC), though its Discovery program.}
 
}
\maketitle
\IEEEpeerreviewmaketitle

 \linespread{1}

\maketitle

\begin{abstract}

When considering the future generation wireless networks, non-orthogonal multiple access (NOMA) represents a viable multiple access technique for improving the spectral efficiency. The basic performance of  NOMA is often enhanced using downlink beamforming and power allocation techniques.  Although  downlink beamforming has   been previously studied with different performance criteria, such as sum-rate and max-min rate,  it has not  been studied in the multiuser,  multiple-input single-output (MISO) case, particularly with the energy efficiency criteria. In this paper, we investigate the design of an  energy efficient beamforming technique for downlink transmission in the context of a multiuser MISO-NOMA system. In particular, this beamforming design is formulated as a global energy efficiency
(GEE) maximization problem with minimum user rate requirements and transmit power constraints. By using the sequential convex approximation (SCA) technique and  the Dinkelbach's algorithm to handle the non-convex nature of the GEE-Max problem, we propose two novel algorithms for solving the downlink beamforming problem for the MISO-NOMA system.  Our evaluation of the proposed algorithms  shows that they offer similar optimal designs and are effective in offering substantial energy efficiencies compared to the designs based on conventional methods.

\end{abstract}

\section{Introduction}
\label{sec:intro}

 Non-orthogonal multiple access (NOMA) is considered to be a viable and a promising multiple access technique to improve the spectral efficiency {  (SE)} in future wireless networks~\cite{fnoma,noma4}. More specifically, it is hailed as an avenue to offer an array of benefits including high spectral efficiency, better quality-of-service (QoS), lower latency and massive connectivity~\cite{noma99,noma2,noma3}. In contrast to   the conventional orthogonal multiple access (OMA) technique,   NOMA simultaneously sends  signals to multiple users  using the {\em power-domain multiplexing}, while sharing  the same time-frequency resources \cite{cuma2}. When performing the power-domain multiplexing {  \cite{ref4},}  multiple signals intended for different users are multiplexed  based on  superposition coding~\cite{noma3} with different power levels prior to the transmission. At the receiving end, the successive interference cancellation (SIC) technique is employed to decode the incoming signals \cite{ding2016}. Furthermore, the massive connectivity offered by NOMA is perfect for handling the requirements of applications stemming from the  area of Internet-of-Things~(IoT), where the transmission of very large number of small messages is rather prevalent~\cite{iot}. In particular,  NOMA can  also be incorporated into a number of future key distributive technologies, such as massive multi-input multi-output (MIMO) systems and { millimeter-wave (mmWave)} technologies {\cite{ref1} \cite{ref3}}. The key drive in such applications is to further increase the throughput, particularly in  the fifth generation (5G) and beyond wireless networks~ \cite{ding2015,mimonom2,crnoma}.

With the progressive adoption of 5G and beyond wireless networks, {     one of the main goals is to achieve a higher spectral efficiency compared to the ones available in contemporary wireless communications systems. Higher spectral efficiency will enable applications that demand  different high data rates and  will provide massive connectivity for IoT~\cite{book}. With limited available wireless resources, including radio spectrum and transmit power, meeting higher data requirements will  only be possible through novel techniques and efficient resource utilizations~\cite{green}. Furthermore, the transmit power required to meet the corresponding throughput requirements with the conventional approaches will be significantly high. This increased power consumption will subsequently induce further issues  such as extra CO$_2$ emission and associated  climate changes \cite{ee2}. Recently, energy efficient techniques  are considered as one of the key avenues for addressing these issues  in the development of future wireless systems \cite{green}.
  The energy efficient designs based on global energy efficiency (GEE) performance metric have become one of the key requirements in the development of future wireless systems. These designs take the energy efficiency (EE) performance metric into account rather than the achievable rate or transmission power metrics. The GEE performance metric  is defined as the ratio between the achievable sum-rate and total power consumption~\cite{ee1,ee2}. Furthermore, the  GEE design can be viewed as a  multi-objective design problem, which aims to simultaneously optimize two conflicting performance metrics, namely, the sum rate and the required power to achieve this sum rate~\cite{ee1}. Furthermore, this performance metric efficiently utilizes the available transmit power while striking a good balance between the achievable sum rate and power consumption. Finally,  unlike  the conventional sum rate maximization  and power minimization designs, the GEE design incorporates the power losses at the base station as part of the design process~\cite{ee2}.
  The EE performance metric plays a crucial role in the overall performance of a NOMA system due to the fact that it exploits power-domain multiplexing to simultaneously transmit signals to multiple users~\cite{noma2}.  Achieving the massive connectivity  offered by NOMA requires a huge amount of transmit power which would  only be possible by considering energy efficient designs \cite{noma3}. Therefore,   the EE performance metric attracts a great deal of attention from the community, which thrives to develop various novel yet practical techniques including NOMA for future wireless networks.}

In the context of NOMA, the core component, the power domain multiplexing,  directly dictates the power allocation at the transmitter-level, and thus indirectly controls the overall energy consumption. Hence, the power domain multiplexing, when combined with   downlink beamforming, will decide the  overall energy performance of a NOMA system. Therefore, considering all the parameters that affect the resulting EE is very crucial for improving the EE of the entire system, particularly at the design level. A number of approaches have been proposed in the literature for this purpose. In~\cite{hanif}, a beamforming design to maximize the spectral efficiency is proposed for multiple-input single-output 
(MISO)-NOMA systems using a variant of the minorization-maximization algorithm~\cite{mm}. In~\cite{misonoma}, a power minimization problem is considered in conjunction with the QoS aspect, and an approach based on the second order cone (SOC) programming is proposed.  An energy efficient-based power allocation scheme is developed in~\cite{EESSISO} for a single-input single-output (SISO) NOMA system by utilizing the Karush-Kuhn-Tucker (KKT) conditions~\cite{convex}. These approaches are complemented by~\cite{ener2}, where both the EE and the computational complexity  arising out of the SIC operations are addressed. They have proposed a clustering method for maximizing the EE for SISO-NOMA systems, with the approach of assigning the same  channel for multiple users. In~\cite{zhu2017optimal}, a max-min fairness based energy efficient design is studied in the context of  downlink transmission for  SISO-NOMA systems. On the other hand, the authors in \cite{zeng2018energy} consider the EE maximization problem for MIMO-NOMA systems, with multiple users are grouped in a cluster. Whereas, multiple resource allocations strategies and clustering algorithms have been introduced in \cite{islam2018} {  \cite{ref2}.} Furthermore, the authors in \cite{fang2} investigated an energy efficiency  design for a downlink
NOMA single-cell network assuming  imperfect channel state information.

To-date, to the best of our knowledge, no study has been conducted on downlink beamforming design for multiuser MISO-NOMA systems, particularly with the focus of maximizing the EE. In this paper, we address this issue, by  proposing an energy-efficient downlink beamforming design.  We formulate the overall problem as a global energy efficiency maximization (GEE-Max) problem that incorporates the minimum rate requirements and transmit power constraints as part of the formulation.\footnote{In this paper, both the GEE and EE carry the same meaning.} However, as will be seen, due to the non-convex nature of the constraints and because of the overall fractional objective function, the overall GEE-Max formulation is a non-convex optimization problem. With the appropriate and necessary initial evaluations, and through suitable approximations, we overcome the non-convexity issues for solving the GEE-Max problem. We make the following key contributions:

\begin{enumerate}
 \item {\bf Meta-approach for detecting the feasibility of solving GEE-Max}: Although the overarching problem can be formulated as a GEE-Max problem, the formulation does not guarantee the solvability of  the GEE-Max problem.  In fact, it might turn out to be an infeasible problem. We propose an approach to detect this infeasibility at the early stages of problem formulation,  and propose an alternative approach;
 \item {\bf Algorithms for solving the GEE-Max problem}:  As discussed above, the non-convexity of the GEE-Max problem stems from two aspects: the non-convex nature of the constraints and the fractional nature of the overall objective function. We present two novel iterative algorithms for handling these issues. The first algorithm uses the sequential convex approximation (SCA) technique to approximate the non-convex constraints while the  second algorithm utilizes the   Dinkelbach's  technique for handling the  non-convex  nature of the fractional objective function; and
 \item {\bf Optimality validation}:  We validate the optimality of the proposed  SCA-based GEE-Max algorithm by comparing the optimal solution of an equivalent power minimization problem. In particular, both the algorithm and power minimization problem  should lead to identical solutions. However, the power minimization problem is a convex problem, and therefore, the solution is optimal in terms of required transmit power to achieve the corresponding minimum rate at each user. This confirms the optimality of the results   obtained through the SCA-based algorithm.
\end{enumerate}

The remainder of the paper is organized as follows.  In Section~\ref{sec:problem}  we present  a system model and formulate the problem with necessary feasibility conditions. This is then followed by Section~\ref{sec:algorithms}, where we propose two iterative algorithms, a key part of our contributions. Detailed  evaluations and simulation results are presented in Section~\ref{sec:evaluation}, demonstrating the effectiveness of the proposed algorithms through comparing their performance with different beamforming techniques available in the literature. We then finally conclude the paper in Section~\ref{sec:conclusions}.

In the rest of the paper, we adopt the following notations: We use the lower-case boldface letters for vectors and upper-case  boldface letters for matrices; $(\cdot )^H $   denotes complex conjugate transpose;  $\Re(\cdot)$ and $\Im(\cdot)$  stand for  real and imaginary parts of a complex number, respectively;  $\mathbb{C}^{N}$ and $\mathbb{R}^{N}$ denote the $N$-dimensional  complex  and real spaces, respectively; and   E($\cdot$),  $||\cdot||_2 $  and $| \cdot| $ represent the expectation, the Euclidean norm of  a vector, and absolute value of a complex number, respectively.   Tr$(\cdot )$ stands for the trace of a matrix.

\section{System Model and Problem Formulation}
\label{sec:problem}

\subsection{System Model}
\begin{figure}[t]
	\centering
	\includegraphics[width=.5\textwidth]{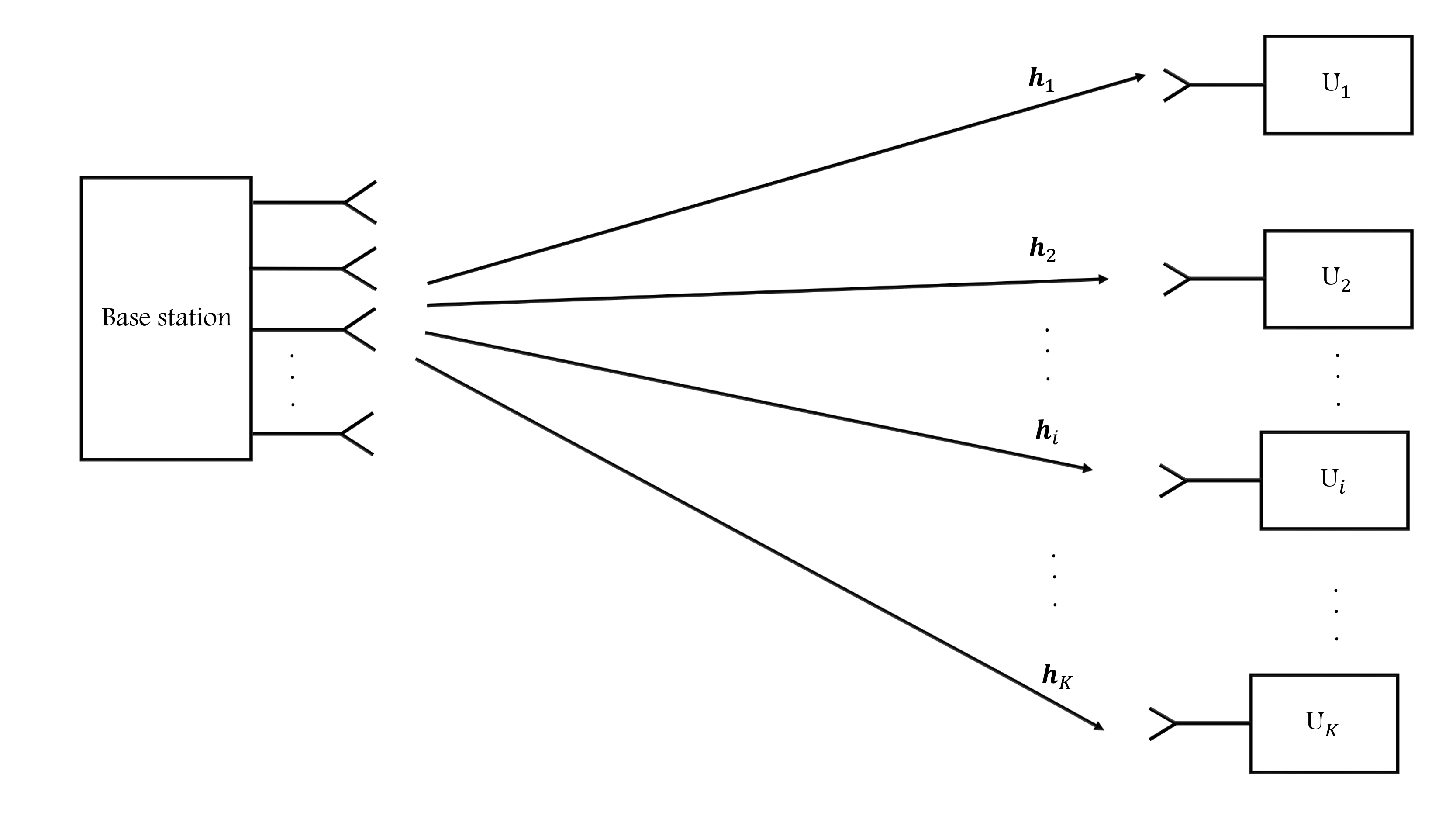}
	\caption{A multiuser, MISO-NOMA system with a multi-antenna base station, and $K$-single antenna users.}
	\label{fig:mumisos1}
\end{figure}

We consider a downlink transmission of a MISO-NOMA system  with $K$ users  as shown in Figure~\ref{fig:mumisos1}, where a  single base station simultaneously transmits information to all users. We assume that the base station here is equipped with $N$ antennas {(i.e., $N>1$)} while there is  a single antenna at the user's end.  The signal transmitted  from the base station can be expressed as:

\begin{equation}
\mathbf{x}=\sum_{k=1}^{K} \mathbf{w}_ks_k,
\end{equation}

\noindent where $s_k $ and  $\mathbf{w}_k  \in \mathbb{C}^{N\times 1}$ denote the signal intended for the $k^{th}$ user and the corresponding beamforming vector, respectively. { Note that    a digital beamforming is adopted in this work.} The power of the transmitted symbol is assumed to be one, i.e., E($|s_{k}|^2) = 1$.

The received signal at the $i^{th}$ user can be expressed  as:

\begin{equation}
y_i=\mathbf{h}_i^H\mathbf{w}_is_i+\sum_{j=1,j\neq i }^{K} \mathbf{h}_i^H \mathbf{w}_j s_j +n_i,
\label{eqn:received_signal}
\end{equation}

\noindent where  $\mathbf{h}_i\in\mathbb{C}^{N\times1}$ represents the channel coefficients between the base station and the $i^{th}$ user, and $n_{i}$ represents the zero-mean circularly symmetric complex additive white Gaussian noise with variance $\sigma^{2}$. Furthermore, we define $\mathbf{h}_i$ such that $\mathbf{h}_i~=~\sqrt{d_i^{-\kappa}}\mathbf{g}_i$, where $\kappa, d_{i}$ and $ \mathbf{g}_{i} $ denote the path loss exponent, the distance between the $i^{th}$ user and the base station, and small scale fading, respectively. { In addition, it is   assumed that the base station has the perfect channel  state information (CSI) of each user}.   In NOMA, user ordering plays a crucial role in implementing the SIC at the users' ends,  and in fact, determines the overall performance of the system  \cite{cuma2}. However, determining optimal user ordering is an NP-hard problem, which can only be solved through exhaustive search,  branch and bound methods or heuristic approaches~\cite{noma2,cumanan}.  In this paper, for the reasons of simplicity, the users are ordered based on their respective channel  strengths. Thus, the first user will have the strongest channel strength, while the channel strength of the $K^{th}$ user will be the weakest. As such, the channels can equivalently be ordered as follows:

\begin{equation}
\LTwoNormSQ{\mathbf{h}_K} \leq \LTwoNormSQ{\mathbf{h}_{K-1}} \leq \cdots \leq \LTwoNormSQ{\mathbf{h}_1}.
\end{equation}
By employing SIC, the $i ^{th}$ user ($U_{i}$) should be able to successively decode and subtract the signals intended for the weaker users $U_{K},\cdots,U_{i-1}$  from the received signal   \cite{cuma9}. The received signal after eliminating the last $K-i$ user signals can expressed as:
\begin{equation}
\overset\sim{y_i}=\mathbf{h}_i^H\mathbf{w}_is_i+\sum_{j=1}^{i-1} \mathbf{h}_i^H \mathbf{w}_j s_j +n_i,
\label{eqn:sara}
\end{equation}

\noindent where the first term in  (\ref{eqn:sara}) represents the intended signal for the $i^{ th}$ user, while the second term denotes the interference caused by the first $i-1$ signals intended for  users  $\{U_1,\cdots,U_{i-1}\}$. {Note that   $U_i$ perfectly  decodes the messages intended to the weaker users  without any errors}.  The achievable signal to interference and noise ratio (SINR) for the $k^{th}$ user to decode the signal that is intended to the $i^{ th}$ user is:

\begin{equation}
SINR_k^i= \frac{| \mathbf{h}_k^H\mathbf{w}_i |^2}{\sum_{j=1}^{i-1}| \mathbf{h}_k^H \mathbf{w}_j |^2+\sigma_k^2}, i\in\mathcal{K}, \forall k\leq i.
\label{eqn:sin2r}
\end{equation}
\noindent It is inherently clear that   (\ref{eqn:sin2r}) holds true only after performing   SIC on the preceding $k-i$ signals. Furthermore, $\mathcal{K} \overset{\bigtriangleup}{=}\{1,2,\cdots,K\}.$ To successfully decode the $i^{th}$ signal at the $k$ strong users (i.e., $k<i$), the achievable SINR of decoding the $i^{th}$ signal at the $k^{th}$ user, $\gamma_k^i$,   should be greater than a predefined threshold $\gamma_{th}$. This condition can be mathematically defined as in~\cite{hanif} \cite{wcnc1}. Thus,

\begin{equation}
 {\gamma}_i= \min  ({\gamma^i_{1},\gamma^i_{2},\cdots,\gamma^i_{k},\cdots,\gamma^i_{i}})\geq\gamma_{th},
\label{eqn:sinr}
\end{equation}

\noindent therefore, the achievable rate $R_i$ for the user  $U_i$ can be defined as follows:

\begin{equation}
R_i=B_w\log_2(1+ {\gamma}_i), \forall i\in\mathcal{K},
\end{equation}
where $B_w$ is the available bandwidth. For notational simplicity, we assume that $B_w=1$  in the rest of this paper. 
Furthermore, the user with the weakest channel strength may not be able to achieve a reasonable rate, owing to the fact that  significant amount of transmit power must have already been  allocated. To circumvent this problem, the following condition  should also be satisfied to maintain a rate fairness between users with different channel strengths~\cite{wcnc2}:

\begin{equation}
 | \mathbf{h}_i^H\mathbf{w}_K |^2 \geq | \mathbf{h}_i^H\mathbf{w}_{K-1} |^2 \geq\cdots\geq| \mathbf{h}_i^H \mathbf{w}_{1}|^2,
 \label{eqn:channels}
\end{equation}

\noindent where $i \in \mathcal{K}$. The constraint expressed in~(\ref{eqn:channels}) will be referred to as the SIC constraint in the rest of this paper.

\subsection{Problem Formulation}

For the  MISO-NOMA system defined above, we consider an  energy efficient maximization problem. In particular, this energy efficient optimization problem is formulated to maximize the  GEE  of the system  satisfying the available total power at the base station, and the minimum rate requirement of each user. The total power consumption at the base station accounts for  both the transmit power allocated for data transmission and the power losses.
As for the required transmit power $P_{tr}$, it  should satisfy the available power budget, $P_{ava}$, at the base station, which  can be mathematically formulated as the following constraint:

\begin{equation}
P_{tr}=\sum_{i=1}^{K}\LTwoNormSQ{\mathbf{w}_i} \leq P_{ava}.
\end{equation}
As for the power losses at the  base station, denoted by $P_{loss}$, they should account both the dynamic and static power losses $p_{dyn}$ and $p_{sta}$, respectively. The former   primarily depends on the number of transmit antennas $N$, whereas the latter   is intended to account for the power required to maintain the system, such as through cooling and conditioning.  The total power losses can be defined as in~\cite{ee1}:

\begin{equation}
P_{loss}=p_{sta}+N p_{dyn}.
\end{equation}
 Hence, the total power consumption at the base station is:

\begin{equation}
P_{total}=\frac{1}{\epsilon_0}P_{tr} + P_{loss},
\end{equation}
  where  $0<\epsilon_0\leq 1$  is the efficiency of the   power amplifier.

The GEE  of the system is defined as the ratio between the total achievable sum rate and the total power consumption,

\begin{equation}
 GEE=\frac{ \sum_{i=1}^{K}R_i}{P_{total}}.
\end{equation}

With these definitions in place, the beamforming design to maximize the GEE in a MISO-NOMA system with $K$ users can be formulated into the following optimization framework:

\bigskip

\begin{subequations}\label{eqn:ee12}
\begin{align}
 { {OP}_1\text{:}~ }
 &\underset{\{\mathbf{w}_i\}_{i=1}^K }{\text{maximize }}
& & \!\!\!\!\!\! \frac { \sum_{i=1}^{K}\log{(1+\gamma_i)}}{\frac{1}{\epsilon_0}\sum_{i=1}^{K}\vert \vert \mathbf{w}_i \vert \vert ^2_2 + P_{loss}  } \label{eqn:frac} \\
&  \text{subject to}
& & \!\!\!\!\!\! R_i \geq R^{\text{min}}_i, \, \forall i \in \mathcal{K},\label{eqn:minrate}\\
       & &&\!\!\!\!\!\!  \sum_{i=1}^{K}\vert \vert \mathbf{w}_i \vert \vert ^2_2\leq P_{ava}, \label{eqn:sa77ar}\\
      & &&\!\!\!\!\!\!(\ref{eqn:channels})   , \label{eqn:sa7ar}
\end{align}
\end{subequations}

\noindent where $R_{i}^{min}$   is the  minimum rate requirement for the user $U_i$, and~(\ref{eqn:sa7ar}) ensures the successful implementation of the SIC for all users while maintaining the rate fairness between them.

\subsection{Feasibility Conditions}

The optimization problem $OP_1$  defined in~(\ref{eqn:ee12}) is worth solving only  when it is  feasible to solve for a given set of constraints. For instance, the $OP_1$ problem may not be solvable because of insufficient available power budget at the base station, or higher user data rate requirements. As such, it is first worth verifying the feasibility conditions prior to attempting to solve the GEE-Max problem. We outline an approach for verifying  the feasibility conditions using the power minimization   (P-Min) problem with minimum data rate requirements and user rate fairness constraints:
\bigskip

\begin{subequations}\label{eqn:power min}
\begin{align}
 {OP_2}\text{:}~P^*\text{=}~
& \underset{\{\mathbf{w}_i\}_{i=1}^K}{\text{minimize }}
& &\!\!\!\!\!\! \sum_{i=1}^{K}\vert \vert \mathbf{w}_i \vert \vert ^2_2  \\
& \text{subject to}
& & \!\!\!\!\!\! R_i \geq R^{\text{min}}_i, \forall i \in \mathcal{K},\label{ghhg}\\
      & && \!\!\!\!\!\! (\ref{eqn:channels}),
\end{align}
\end{subequations}

\noindent where $P^*$ denotes the minimum power required to achieve the minimum rate and satisfy the SIC constraints. The above optimization problem,  $OP_2$, has been  solved in~\cite{misonoma} by handling the non-convex constraints through  a set of convex approximation techniques, which are detailed  in the next section. If $P^*>P_{ava}$, then the optimization problem in $OP_1$ can be classified as an infeasible problem. To overcome this infeasibility issue, we consider the following sum-rate maximization (SRM) problem, where the achievable sum-rate is maximized with transmit power constraint and SIC constraints. This SRM problem can be formulated   as in~\cite{hanif}:

\bigskip

\begin{subequations}\label{srm}
\begin{align}
  {OP_3}\text{:}~
  &\underset{\{\mathbf{w}_i\}_{i=1}^K}{\text{maximize }}
& &  \!\!\!\!\!\! { \sum_{i=1}^{K}\log{(1+\gamma_i)}} \label{tay} \\
& \text{subject to}
 &&  \!\!\!\!\!\!  \sum_{i=1}^{K}\vert \vert \mathbf{w}_i \vert \vert ^2_2\leq P_{ava},\\
      & &&  \!\!\!\!\!\!    (\ref{eqn:channels}).
\end{align}
\end{subequations}

The optimization problem $OP_3$ is solved as in~\cite{hanif} using  the SCA technique. It is worth to explore an efficient method to solve the original GEE-Max problem in $OP_1$, provided that the problem is feasible. In the following section, we develop two iterative algorithms to determine a solution with the assumption that the minimum data requirements and the SIC constraints can be met within the available power budget.

\section{Algorithms for Solving the GEE-Max Problem}
\label{sec:algorithms}

The GEE-Max problem defined in  (\ref{eqn:ee12}) is a non-convex problem due to non-convex objective function and constraints. Hence, it is challenging to obtain an optimal solution. In this section, we develop two iterative algorithms to determine the beamforming vectors to maximize the GEE of the system while satisfying the respective constraints. These algorithms are proposed by approximating the non-convex objective function and constraints to   convex ones based on the SCA  and Dinkelbach's algorithms. The details of the algorithms are provided in the following two subsections.


\subsection{Approach based on the Sequential Convex Approximation}

Sequential convex approximation or sequential convex programming~\cite{sqa}  is one of the well-known techniques that has been widely adopted to approximate and transform non-convex problems  to  convex problems. A number of  studies demonstrate the viability of this approach in real-world applications~\cite{hanif,ee2}. The basic idea of SCA is to establish a convex trust region around the original non-convex spatial points, so that the overall objective function is sequentially optimized for every spatial point. As such, it is a heuristic-driven approach, where the optimality of the ultimate solution may vary depending on the initialization. In our case, by introducing a slack variable $\alpha$, the original problem specified by $OP_1$ can be reformulated in the following epigraph form:

\begin{subequations}\label{eqn:e77}
\begin{align}
&    \underset{\{\mathbf{w}_i\}_{i=1}^K}{\text{maximize }}
&  \!\!\!\!\!\!  \alpha \label{eqn:frac7} \\
& \text{subject to}
& &  \!\!\!\!\!\! \frac {\sum_{i=1}^{K}\log{(1+\gamma)}}{\frac{1}{\epsilon_0}\sum_{i=1}^{K}\vert \vert \mathbf{w}_i \vert \vert ^2_2 + P_{loss}  } \geq \sqrt{\alpha} , \label{eqn:rar}\\
& &&  \!\!\!\!\!\! R_i \geq R^{\text{min}}_i, \,   \forall i \in \mathcal{K},\label{eqn:minrate7}\\
       & &&  \!\!\!\!\!\!  \sum_{i=1}^{K}\vert \vert \mathbf{w}_i \vert \vert ^2_2\leq P_{ava}, \label{eqn:sa77ar7}\\
      & &&   \!\!\!\!\!\!  (\ref{eqn:channels}), \label{eqn:sa7ar7}
\end{align}
\end{subequations}

\noindent where the objective function in the original problem  (\ref{eqn:ee12}) is replaced by  $ \sqrt{\alpha}$ (or equivalently ${\alpha}$). Without loss of generality, the  non-convex constraint with the slack variable $\alpha$ in~(\ref{eqn:rar}) can  equivalently be decomposed  into the following two constraints:

\begin{subequations}
\begin{align}
\sum_{i=1}^{K}\log{(1+\gamma_i)} \geq \sqrt{\alpha \beta},
\label{eqn:eqn11}\\
\frac{1}{\epsilon_0}\sum_{i=1}^{K} \LTwoNormSQ{\mathbf{w}_i} + P_{loss} \leq \sqrt{\beta}.
\label{eqn:eqn12}
\end{align}
\end{subequations}

By incorporating the definition of SINR$_i$ (i.e., $\gamma_i$)   in  (\ref{eqn:sinr}),  the constraint in  (\ref{eqn:eqn11}) can be represented as follows:

\begin{equation}
\label{eqn:sdsdsd}
\sum_{i=1}^{K}\log{(1+\min  ({\gamma^i_{1},\gamma^i_{2},\cdots,\gamma^i_{k},\cdots,\gamma^i_{i}}))}\geq \sqrt{\alpha \beta}.
\end{equation}

To handle the non-convexity of (\ref{eqn:sdsdsd}), we firstly introduce a set of  new slack variables    such that:

\begin{subequations}
\begin{align}
\log{(1+  \min  ({\gamma^i_{1},\gamma^i_{2},\cdots,\gamma^i_{k},\cdots,\gamma^i_{i}}))}\geq \delta_i , \forall i \in \mathcal{K},\label{eqn:min}\\
(1+{\min  ({\gamma^i_{1},\gamma^i_{2},\cdots,\gamma^i_{k},\cdots,\gamma^i_{i}})})\geq   \zeta_i,\forall i \in \mathcal{K}.\label{eqn:asad}
\end{align}
\end{subequations}

\noindent   Based on these new slack variables, the constraint in~(\ref{eqn:min}) can  equivalently be represented by the following set of  constraints:

 \begin{subnumcases}{
 (\ref{eqn:min})  \Leftrightarrow  }
    \sum_{i=1}^{K} \delta_i \geq  \sqrt{\alpha \beta},
     \label{eqn:paart}\\
   \zeta_i \geq 2^{\delta_i}, \quad \forall i \in \mathcal{K}.
   \label{eqn:part2}
\end{subnumcases}

However, the constraint in~(\ref{eqn:paart})   still remains   non-convex. In order to relax this, we exploit the first-order Taylor  series,  providing approximations around  the values of  $ (\alpha^{(n-1)},\beta_{k,i}^{(n-1)})$. With this,


\begin{multline}
  \sum_{i=1}^{K} \delta_i \geq  \sqrt{\alpha^{(n-1)} \beta^{(n-1)}} 
 + \\0.5\sqrt{\frac{\alpha^{(n-1)}}{ \beta^{(n-1)}}}(\beta^{(n) }-\beta^{(n-1)})  
    \\+ 0.5\sqrt{\frac{\beta^{(n-1)}}{ \alpha^{(n-1)}}}(\alpha^{(n) }-\alpha^{(n-1)}).\label{eqn:joodeh}
\end{multline}

To handle  the constraint expressed in~(\ref{eqn:asad}), we introduce another  slack variable, $\theta_{k,i} \in \mathbb{R}^{1}_+$ ($\forall i\in \mathcal{K}, k\leq i$ ),  and reformulate the constraint in   (\ref{eqn:asad}) into the following set of equivalent constraints, with $\forall i\in \mathcal{K}$ and  $k\leq i$,

\begin{subequations}
\label{eqn:concom}
\begin{align}
\vert \mathbf{h}_k^H\mathbf{w}_i  \vert^2\geq (\zeta_i-1)\theta_{k,i},
 \label{eqn:hkh}\\
\sum_{j=1 }^{i-1} \vert \mathbf{h}_k^H \mathbf{w}_j  \vert^2+\sigma_i^2\leq {\theta_{k,i}}.
\label{eqn:hskh}
\end{align}
\end{subequations}

Without any loss of generality, the constraint in~(\ref{eqn:hkh}), can be re-expressed  by introducing an arbitrary rotation to the phase  of the beamforming vector  $\mathbf{w}_i$  such that  $ \mathbf{h}_k^H\mathbf{w}_i  $ becomes real, while the imaginary part $\Im(\mathbf{h}_{k}^{H}\mathbf{w}_{i})$ becomes zero. This change, however, does not alter the original optimization problem nor the solution owing to the fact that this change in the beamformer does not  modify the required transmit power or the achieved SINR of any users~\cite{cvx2}. Therefore, the constraint covered by~(\ref{eqn:hkh}) can  equivalently be expressed as follows:

\begin{equation}
\label{eqn:ta}
\Re{( \mathbf{h}_k^H\mathbf{w}_i )}\geq  \sqrt{({\zeta_i-1)}\theta_{k,i}},
\end{equation}

\noindent where the right side of the above inequality can be approximated using the first-order Taylor series, and the inequality in~(\ref{eqn:ta}) becomes:

\begin{multline}
\label{eqn:tata}
 {\Re (\mathbf{h}_k^H\mathbf{w}^{(n) }_i )}\geq \sqrt{{(\zeta_i^{(n-1)}-1)} \theta_{k,i}^{(n-1)}}   \\  
  +0.5 \sqrt{\frac{({\zeta_i^{(n-1)}-1)}}{\theta_{k,i}^{(n-1)}}}(\theta_{k,i}^{(n) }-\theta_{k,i}^{(n-1)})\\
 +0.5 \sqrt{\frac{\theta_{k,i}^{(n-1)}}{{(\zeta_i^{(n-1)}-1)}}}(\zeta_i^{(n) }-\zeta_i^{(n-1)}).
\end{multline}

On the other hand, the non-convex constraint in  (\ref{eqn:hskh}) can be formulated
into a SOC as in~\cite{fayzeh}:

\begin{equation}
\label{eqn:tata1}
\begin{split}
\frac{\theta_{k,i}^{(n)}-\sigma_i^2+1}{2}
 & \geq {||{\mathbf{v}_k^{(n)}}||}_2,
\end{split}
\end{equation}

\noindent where

\begin{equation}
  \mathbf{v}_k^{(n)} =
  \Bigl[
    \mathbf{h}_k^H \mathbf{w}^{(n)}_{1}   \ldots   \mathbf{h}_k^H \mathbf{w}^{(n)}_{i-1} ~ \phi_{i,k} 
  \Bigr]^T
\end{equation}

\noindent and

\begin{equation}
\phi_{i,k}^n = \frac{(\theta_{k,i}^{(n) }-\sigma_i^2)-1}{2}.
\end{equation}

 The non-convexity of the constraint in  (\ref{eqn:eqn12}) can be handled by introducing new slack variable $ \beta_{\beta}$ and expressed into multiple, yet equivalent,  constraints as follows:

\begin{subequations}
\begin{align}
& \sqrt{\beta}\geq \beta_{\beta},
\label{eqn:wq}\\
&\beta_{\beta} \geq\frac{1}{\epsilon_0}\sum_{i=1}^{K}\LTwoNormSQ{ \mathbf{w}_i } + P_{loss}.
 \label{eqn:wq1}
\end{align}
\end{subequations}

This can  further be formulated into the following SOC  constraints~\cite{convex}:

\begin{subequations}
\begin{align}
 \frac{\beta+1}{2} &\geq \BigLTwoNorm{\Bigl[\frac{\beta-1}{2}  \quad \beta_{\beta}\Bigr]^T},
 \label{eqn:wssdq}\\
 \frac{(\beta_{\beta}-{P_{loss}})+1}{2} &\geq \BigLTwoNorm{\Bigl[
 w_0 ~ \frac{\mathbf{w}_1}{\sqrt{\epsilon_0}}  \ldots   \frac{\mathbf{w}_K}{\sqrt{\epsilon_0}}
 \Bigr]^T
 },\label{eqn:wq1s} 
\end{align}
\end{subequations}

\noindent where:

\begin{equation}
w_0 =   \frac{(\beta_{\beta}-{P_{loss}})-1}{2}.
\end{equation}

Having approximated the original non-convex objective function in~(\ref{eqn:frac})   by introducing a number of slack variables, the final form is such that the problem is same as:

\begin{subnumcases}{(\ref{eqn:frac})  \Leftrightarrow  }
   \text{maximize}\quad \alpha\nonumber
    \label{eqn:padsdart}
   \\
  \text{subject to}
   ~ (\ref{eqn:part2}),(\ref{eqn:joodeh}),(\ref{eqn:tata}),\nonumber\\
\quad  \quad \quad \quad \quad(\ref{eqn:tata1}),(\ref{eqn:wssdq}),(\ref{eqn:wq1s})\nonumber.
  \label{eqn:pasdsdrt2}
\end{subnumcases}

 With this,  we handle other non-convex constraints in the original GEE-Max problem expressed by $OP_{1}$. Without the loss of generality, the minimum rate constraint in  (\ref{eqn:minrate}) can be expressed as:

\begin{equation}
  \label{eqn:ghjsd}
\gamma^i_k\geq \gamma_i^{min} ,\forall i\in \mathcal{K}, k\leq i,
\end{equation}

\noindent where $\gamma_{i}^{min} = 2^{R_{i}^{min}} - 1$. Furthermore, this SINR constraint can be formulated into an SOC by introducing an arbitrary phase rotation as in~(\ref{eqn:hkh}) and with $\forall i\in \mathcal{K}, k\leq i$

\begin{equation}
  \label{eqn:const}
\frac{1}{\sqrt{\gamma_i^{min}}} \Re( \mathbf{h}_k^H\mathbf{w}_i )\geq \BigLTwoNorm{\Bigl[ \mathbf{h}_k^H \mathbf{w}_{1}  \ldots \mathbf{h}_k^H \mathbf{w}_{i-1}~\sigma_k\Bigr]^T}.
\end{equation}

The  non-convexity of the constraint in~(\ref{eqn:sa7ar}) can be approximated to a convex constraint by applying the first-order Taylor series approximation. However, instead of applying the Taylor series expansion to the original equation,  we define a new proxy function $f(\psi_{i,j})$ by stacking the real and imaginary parts of the product $\mathbf{h}_i^H\mathbf{w}_j $ as follows:

\begin{eqnarray}
f(\psi_{i,j}) &=& \LOneNormSQ{\mathbf{h}_i^H\mathbf{w}_j} \nonumber\\
             & = &  |\LOneNormSQ{[\Re( \mathbf{h}_i^H\mathbf{w}_j) ~ \Im({ \mathbf{h}_i^H\mathbf{w}_j})]^T|},
\end{eqnarray}

\noindent where $ \psi_{i,j}=[\Re( \mathbf{h}_i^H\mathbf{w}_j ) ~\Im({ \mathbf{h}_i^H\mathbf{w}_j })]$. We then apply the first-order Taylor series expansion to this proxy function $f(\psi_{i,j})$, and consider two most significant terms:

\begin{equation}
f(\psi_{i,j}^{(n) })\cong f(\psi_{i,j}^{(n-1)})+2(\psi_{i,j}^{(n-1)})^T(\psi_{i,j}^{(n) }-\psi_{i,j}^{(n-1)}).
\end{equation}

With this  approximation in place, each element in the inequality constraint in~(\ref{eqn:sa7ar}) can be replaced by the following linear function:

\begin{multline}
  \label{eqn:anoora}
   \VLTwoNormSQ{\mathbf{h}_k^H\mathbf{w}_j^{(n)} } \cong \\
   \VLTwoNormSQ{
  \begin{bmatrix}
          {r}_{k,j}^{(n-1)} \\
          {i}_{k,j}^{(n-1)}
  \end{bmatrix}
  }   
  + 2
 \begin{bmatrix}
        {r}_{k,j}^{(n-1)}  \\
          {i}_{k,j}^{(n-1)}
 \end{bmatrix}
 \begin{bmatrix}
         {r}_{k,j}^{(n )} -  {r}_{k,j}^{(n-1)} \\
         {i}_{k,j}^{(n )} -  {i}_{k,j}^{(n-1)}
 \end{bmatrix}^{T},
\end{multline}

\noindent where

\begin{equation}
  {r}_{k,j}^{(n )}= \Re{(\mathbf{h}_k^H \mathbf{w}_j^{(n)})}, 
\end{equation}

\noindent and

\begin{equation}
   {i}_{k,j}^{(n )} = \Im{(\mathbf{h}_k^H \mathbf{w}_j^{(n)})}.
\end{equation}

By incorporating all of these approximations, the original GEE-Max problem in  (\ref{eqn:ee12})   can be formulated into the following approximated  problem:

\begin{subequations}
  \label{eqn:se12}
\begin{alignat}{4}
&\underset{ \Lambda }{\text{maximize }}
 &&\alpha   \\
 &\text{subject to}
 & &(\ref{eqn:sa7ar}), (\ref{eqn:asad}), (\ref{eqn:tata}),(\ref{eqn:tata1}), (\ref{eqn:const}),  \\
  & && (\ref{eqn:joodeh}),(\ref{eqn:wssdq}),(\ref{eqn:wq1s}), (\ref{eqn:sa77ar}). 
\end{alignat}
\end{subequations}

\noindent Note that we replaced~(\ref{eqn:anoora}) instead of each term in the inequality in~(\ref{eqn:sa7ar}), and $\forall i\in \mathcal{K}, k\leq i$, wherever applicable. Furthermore, the expression

$$\Lambda^{(n)} \overset{\bigtriangleup}{=}\{\mathbf{w}_i^{(n)},\alpha^{(n)},\beta^{(n)},\beta_{\beta}^{(n)},\theta_{k,i}^{(n)},\zeta_i^{(n)},\delta_i^{(n)} \}_{i=1}^{K}
$$

\noindent indicates the $n^{th}$ iteration of the  optimization parameters.  In particular, the original GEE-Max problem will be iteratively solved  using the  approximated convex problem in~(\ref{eqn:se12}). As such, the   optimization parameter  is initialized with  $\Lambda^{(0)}$. In particular,  the random selection of $\Lambda^{(0)}$ determines both the feasibility and the convergence  of~(\ref{eqn:se12}). Hence, we initialize $\Lambda^{(0)}$ by  evaluating the beamforming vectors first that satisfy  the constraints specified by the  optimization problem in~(\ref{eqn:power min}), where the initial slack variables $\alpha^{(0)},\beta^{(0)},\beta_{\beta}^{(0)},\theta_{k,i}^{(0)},\zeta_i^{(0)}$ and $\delta_i^{(0)}$
are found  by substituting these initial beamforming vectors in   (\ref{eqn:eqn11}), (\ref{eqn:eqn12}), (\ref{eqn:wq}),  (\ref{eqn:hskh}),  (\ref{eqn:hkh}), and (\ref{eqn:asad}), respectively.  We summarize the algorithm   developed to determine the solution of the original GEE-Max  problem in  Algorithm~\ref{alg:sca}. The algorithm  is terminated when the absolute difference between two sequential optimal values is less than  a pre-defined   threshold  $\varepsilon $ (i.e., $|\alpha^{(n)}-\alpha^{(n-1)}|< \varepsilon $).

\begin{algorithm}
     \caption{GEE-Max using SCA}
       \label{alg:sca}
      \hspace{0em}\hrulefill\\
     \hspace{0em} \textbf{S}tep 1: Initialization of  $\Lambda^{(0)}$
     
     \hspace{0em} \textbf{S}tep 2: Repeat
\begin{enumerate}
   \item Solve the  optimization problem in~(\ref{eqn:se12}).
   \item Update $ \Lambda^{(n)} $ .
\end{enumerate}
\hspace{-0em} \textbf{S}tep 3: Until required accuracy is achieved.
 
  \hrulefill
\end{algorithm}

\subsection{Approach based on the Dinkelbach's Algorithm}
\label{sec:evaluation:dinkel}

In this subsection, we develop another approach based on the  Dinkelbach's
algorithm to solve the   GEE-Max problem in $OP_{1}$. In addition
to offering an alternative, this approach also helps to compare and
validate the performance of the SCA-based algorithm, and   minimize the number of required
slack variables in  the   algorithm. Furthermore, although the
SCA-based algorithm was useful in transforming the non-convex
constraints into convex ones, the fractional nature of the objective
function in $OP_{1}$ still remains untouched. We address this issue by introducing an
additional non-negative variable, and by representing the objective
function by a parametrized, yet equivalent, non-fractional function. The non-negative variable we introduce here,
$\chi$,   based on~\cite{dink}, is as follows:
\begin{subequations}
  \label{eqn:de12}
\begin{align}
 {OP_5}\text{:}~ {\text{maximize }}
   &\Big( {\sum_{i=1}^{K}\log{(1+\gamma_i)}}-\nonumber \\&\chi \big({\frac{1}{\epsilon_0}\sum_{i=1}^{K}\vert \vert \mathbf{w}_i \vert \vert ^2_2 + P_{loss} \big) }\Big)
\label{eqn:dda} \\
 \text{subject to}~
&    R_i \geq R^{\text{min}}_i, \, \forall i\in \mathcal{K}, k\leq i,\\
       &    \sum_{i=1}^{K}\vert \vert \mathbf{w}_i \vert \vert ^2_2\leq P_{ava},\\
      &      (\ref{eqn:channels}).
\end{align}
\end{subequations}

For the reasons of notational simplicity, we denote the numerator and the denominator of the objective function in $OP_1$  by $f_1$ and $f_2$, respectively, such that

\begin{subequations}
\begin{align}
{f_1(\{\mathbf{w} _i \}_{i=1}^{K} )}= { \sum_{i=1}^{K}\log{(1+\gamma_i)}},\\
{f_2(\{\mathbf{w} _i \}_{i=1}^{K} )}={\frac{1}{\epsilon_0}\sum_{i=1}^{K}\vert \vert \mathbf{w}_i \vert \vert ^2_2 + P_{loss}  }.
\end{align}
\end{subequations}

In order to realize the relationship between $OP_{1}$ and $OP_{5}$, we present the following theorem ~\cite{dink}:

\begin{theorem}
  \label{th:th1}
 A necessary and sufficient condition for

 \begin{equation}
   \label{eqn:Nnn}
 \chi^*=\underset{ \{\mathbf{w} _i \}_{i=1}^{K}}{\text{maximize}}\frac{f_1(\{\mathbf{w} _i \}_{i=1}^{K} )}{f_2(\{\mathbf{w} _i \}_{i=1}^{K} )}=\frac{f_1(\{\mathbf{w}^*_i \}_{i=1}^{K} )}{f_2(\{\mathbf{w}^*_i \}_{i=1}^{K} )},
 \end{equation}

\noindent is

\begin{multline}\label{eqn:param}
  F({\{\mathbf{w}_i } \}_{i=1}^{K},\chi^*)=\\\underset{ \{\mathbf{w} _i \}_{i=1}^{K}}{\text{maximize}}\Big(f_1(\{\mathbf{w} _i \}_{i=1}^{K} )-\chi^*f_2(\{\mathbf{w} _i \}_{i=1}^{K} )\Big)\\=f_1(\{\mathbf{w} _i^* \}_{i=1}^{K} )-\chi^*f_2(\{\mathbf{w} _i^* \}_{i=1}^{K}) =0,
\end{multline}

\noindent where $\{\mathbf{w}^*_i \}_{i=1}^{K}$ are the solution of the original GEE-Max problem.
\end{theorem}

\textit{Proof}: Please refer to  Appendix~A.\\

\smallskip

Theorem~\ref{th:th1} confirms that obtaining the beamforming vectors
that maximize the GEE in the original problem $OP_1$ is the same as
solving the parametrized optimization problem $OP_5$. However, the
precondition is that the non-negative parameter $\chi$ is a solution
of~(\ref{eqn:param}), while $F(\chi) $ has a unique solution
for any set of $\{ \mathbf{w}_i \}_{i=1}^{K}$ as in~\cite{dink}.
In this approach, the design variables $\chi$ and
$\{\mathbf{w}_{i}\}_{i=1}^{K}$ in $OP_{5}$ are iteratively optimized
by exploiting the Dinkelbach's algorithm. First, the parameter $\chi$ is
initialized with zero and the parametrized optimization problem in
 (\ref{eqn:de12}) can be  solved   using  
convex approximation techniques \cite{convex}. For a given set of
beamformers, the design parameter in the $n^{th}$ iteration $\chi$ is
updated as follows:

\begin{equation}
  \label{eqn:taha}
\chi^{(n) }= \frac{f_1(\{{  {\mathbf{w}}}^{(n-1) }_i\}_{i=1}^{K} )}{f_2(\{{ {\mathbf{w}}}^{(n-1) }_i\}_{i=1}^{K} )}.
\end{equation}
In particular, the beamforming vectors in the $n^{th}$ iteration ($
\{\mathbf{w}^{(n) }_i \}_{i=1}^{K} $) can be found by solving the
following optimization problem:
\begin{subequations}
  \label{eqn:ee1s2}
\begin{align}
  \underset{\{\mathbf{w}^{(n) }_i \}_{i=1}^{K} }{\text{maximize }}\Big(
  &{\sum_{i=1}^{K}\log{(1+\gamma_i)}}-\nonumber \\&\chi^{(n-1) }\big({\frac{1}{\epsilon_0}\sum_{i=1}^{K}\vert \vert \mathbf{w}_i^{(n) } \vert \vert ^2_2 + P_{loss} \big) }\Big)
\label{eqn:kawta}  \\
\text{subject to}~
  &  R_i \geq R^{\text{min}}_i,\forall i\in \mathcal{K}, k\leq i,\\
 &  \sum_{i=1}^{K}\vert \vert \mathbf{w}_i^{(n) } \vert \vert ^2_2\leq P_{ava}, \label{eqn:noss}\\
  &    (\ref{eqn:channels}).
\end{align}
\end{subequations}
Due to the non-convex objective function and the non-convex
constraints of the optimization problem in~(\ref{eqn:ee1s2}), we  develop an iterative
algorithm  using the SCA approach. Hence, the
objective function is reformulated to a concave form and it is obvious
that $ \big(\frac{1}{\epsilon_0} \sum_{i=1}^{K}\vert \vert $  $
\mathbf{w}^{(n)}_i\vert \vert ^2_2 + P_{loss} \big) $ is convex.
Hence, without any loss of generality, multiplying it with $-\chi^{(n)}$ will ensure the concavity of this part. On the other
hand, the first part of the  objective function (i.e., $
\sum_{i=1}^{K}\log{(1+\gamma_i)}) $ requires some relaxations to convert
it to a concave form. To this end, we first introduce a new slack
variable such that:

\begin{equation}
  \label{eqn:oko}
\sum_{i=1}^{K}\log{(1+\gamma_i)} \geq \nu,
\end{equation}
where the left side of the inequality in~(\ref{eqn:oko}) can be
approximated by incorporating new slack variables $ {z}_i$ and $
{q}_i$, such that

\begin{subequations}
 \begin{align}
    1+\gamma_i \geq z_i  ,\forall i\in \mathcal{K},
    \label{eqn:mofa} \\
    z_i  \geq 2^{q_i },\forall i\in \mathcal{K}.
    \label{eqn:mofaq}
 \end{align}

Hence, the non-convex constraint in (\ref{eqn:oko}) can be
equivalently rewritten as

\begin{equation}
\sum_{i=1}^{K} { q_i }\geq \nu.
\label{eqn:ommm2}
\end{equation}
\end{subequations}

Without any loss of generality, we can address the non-convexity of
the constraint in (\ref{eqn:mofa}) by following the same approach that has been
developed to approximate the constraint in (\ref{eqn:asad}) in the previous
subsection. In particular, we replace $\zeta_i$ and $\theta_{k,i}$ in~(\ref{eqn:hkh}) and~(\ref{eqn:hskh}) by $z_i$ and  $\rho_{k,i}$, respectively. The constraint in~(\ref{eqn:mofa}) can be equivalently re-written as a set of the following convex constraints:

\begin{multline}\label{eqn:noord}
{\Re (\mathbf{h}_k^H\mathbf{w}_i^{(n) } )}\geq  \sqrt{({z_i^{(n-1)}-1)}\rho_{k,i}^{(n-1)}}  \\+0.5 \sqrt{\frac{({z_i^{(n-1)}-1)}}{\rho_{k,i}^{(n-1)}}}(\rho_{k,i}^{(n )}-\rho_{k,i}^{(n-1)})\\+0.5 \sqrt{\frac{\rho_{k,i}^{(n-1)}}{({z_i^{(n-1)}-1)}}}(z_i^{(n )}-z_i^{(n-1)}), \forall i\in \mathcal{K}, k\leq i.
\end{multline}

\begin{multline}\label{eqn:omi}
\frac{(\rho_{k,i}^{(n )}-\sigma_i^2)+1}{2} \geq  \vert \vert [ \mathbf{h}_k^H \mathbf{w}^{(n )}_{1}~   \mathbf{h}_k^H \mathbf{w}^{(n )}_{2} \cdots    \mathbf{h}_k^H \mathbf{w}^{(n )}_{i-1} \\\frac{(\rho_{k,i}^{(n )}-\sigma_i^2)-1}{2} ] \vert \vert_2, \forall i\in \mathcal{K}, k\leq i.
\end{multline}
So far, we have transformed the first part of the objective function
in $OP_5$ into a concave form. Meanwhile, we will use approaches that
are similar to that of handling the constraints of $OP_1$, as both the
optimization problems have the same constraints. Based on these new
transformations, the GEE-Max problem based on the parametrized objective
function can be expressed as follows:
\begin{subequations}\label{eqn:seeSDs12}
\begin{align}
  \underset{\Upsilon^{(n)}}{\text{maximize }}
&   \nu^{(n)}-\chi^{(n-1)}\big(\frac{1}{\epsilon_0}\sum_{i=1}^{K}\vert \vert \mathbf{w}_i^{(n)} \vert \vert ^2_2 + P_{loss} \big)  \\
 \text{subject to}~\label{eqn:lklk}
& (\ref{eqn:noord}),(\ref{eqn:omi}),(\ref{eqn:const}),(\ref{eqn:sa7ar})  , \forall i\in \mathcal{K}, k\leq i, \\
 &    (\ref{eqn:mofaq}), \forall i\in \mathcal{K},\\
 &  (\ref{eqn:noss}),(\ref{eqn:ommm2}).
\end{align}
\end{subequations}


\noindent   Furthermore, the parameters $\Upsilon$ obtained through
the $n^{th}$ iteration for the new relaxed optimization problem in
 (\ref{eqn:seeSDs12}) are denoted by $\Upsilon^{(n)}$, such as
$\Upsilon^{(n)}=\{\nu^{(n)},{z}_i^{(n)},\rho_{k,i}^{(n)},\mathbf{w}_i^{(n)}\}_{i=1}^{K}$.

In this Dinkelbach's-based iterative algorithm, there are two steps
that have to be carried out to determine the solution of the original
GEE-Max problem, $OP_1$.  These steps involve iteratively determining the optimal
beamforming vectors that would solve the  optimization~(\ref{eqn:seeSDs12})
for different values of $\chi$ until the required accuracy thresholds (i.e., $\varepsilon$ and $\varsigma$) are achieved.   The overall process is outlined in Algorithm~\ref{alg:dinkel}. {  In particular, the parametrization offered by the  Dinkelbach's algorithm might simplify the original problem, especially   when the design parameters are scalar variables, for example,  power allocations, time slots, and bandwidth allocations  \cite{ee1}. In such cases, dealing with non-fractional objective  functions are relatively easier in terms of computational complexity than considering the original fractional objective functions. However,  the parametrization offered by the Dinkelbach's algorithm does not reduce much complexity in our original GEE-Max problem {due to the fact that the parametrized problem   still remains   not-convex, and hence, an additional SCA technique is yet required to handle this non-convexity issue.}   Note that  the Dinkelbach's algorithm is presented as an alternative method to   validate the performance  of    the proposed SCA technique.}

\begin{algorithm}
        \caption{GEE-Max using Dinkelbach's Algorithm.}
        \label{alg:dinkel}      
        \hspace{0em}\hrulefill \\
      \hspace{0em} \textbf{S}tep 1: Initialize $\chi^{(0)} =0$, choose feasible values for $\rho_{k,i}^{(0)} ,\nu^{(0)}$ and $z_i^{0}$.
      
      \hspace{0em} \textbf{S}tep 2: Repeat 
      
       \hspace{0em} \textbf{S}tep 3: Repeat  
\begin{enumerate}
   \item Solve  the optimization problem in (\ref{eqn:seeSDs12}).
   \item    Update $\Upsilon^{(n)}$.
   \end{enumerate}
    \hspace{0em} \textbf{S}tep 4:  Until required accuracy is achieved.
   
\hspace{-0em} \textbf{S}tep 5: Update $\chi ^{(n)} =\frac{\nu^{(n-1)}}{\frac{1}{\epsilon_0}\sum_{i=1}^{K}\vert \vert \mathbf{w}_i ^{(n-1)}\vert \vert ^2_2 + P_{loss}}$.

\hspace{-0em} \textbf{S}tep 6: Until required accuracy is achieved.

\hrulefill

\end{algorithm}
This iterative algorithm for obtaining the solution terminates when the
absolute difference between two consecutive solutions of the parameter
is less than a predefined threshold of $\varepsilon$. Furthermore, we
introduce the following Lemma to confirm the convergence of the
proposed  algorithm.
\begin{lemma}
   \noindent {The GEE-Max using Dinkelbach's Algorithm converges to the   solution after finite iterations}.\\
   \noindent That is
     $$
     \underset{n\rightarrow\infty}{\text{lim}}F(\{\mathbf{w}^{(n) }_i \}_{i=1}^{K},\chi^{(n) }) \to 0.
     $$
   \label{lem:conv-dinkel}
\end{lemma}
\noindent \textit{Proof}: Please refer to Appendix B.

It is worth making two important observations regarding the
solution of the original GEE-Max problem $OP_1$ here. First, unlike
the sum-rate (i.e., $\sum_{i=1}^{K} \log(1+\gamma_i)$) in $OP_3$, GEE is
not monotonically increasing with the available power. However, the
maximum GEE in $OP_1$ is achieved within certain available power
budget, which is referred to as the {\em green power}. In particular,
the GEE remains constant for any available power that is more than the
green power. Secondly, the GEE-Max problem $OP_1$ and SRM problem
$OP_3$ provide similar or same set of solutions (i.e., beamforming
vectors and GEE) for any available power budget that is less than the
green power.

\subsection{Optimality Validation for the GEE-Max Algorithms}

In the previous subsections, the original GEE-Max problem $OP_1$ is
solved through two iterative algorithms, which are developed by
approximating non-convex functions. However,
it is important to validate the optimality of the corresponding
solutions and evaluate their performances by comparing them with the
optimal results. { To validate the optimality of the SCA based algorithm,}  we formulate an equivalent convex
problem and compare the corresponding performance with  the developed
algorithms. In particular, we revisit the P-Min problem in $OP_{2}$
and reformulate it into a semidefinite program (SDP) by relaxing
non-convex rank-one constraint. Furthermore, the achieved SINRs { ($\overset{*}\gamma_i $)}  in the
solution to the original GEE-Max problem $OP_{1}$ are set as the
target SINRs in this P-Min problem.  Without loss of generality, we
introduce new rank-one matrices such that
$\mathbf{W}_i=\mathbf{w}_i\mathbf{w}_i^{H}$ and   reformulate the
problem $OP_{2}$ in the following SDP \cite{orreston} \cite{cuma11}:

\begin{subequations}\label{sdp}
\begin{align}
 {\overset{\sim}{OP_2}}\text{:}~ P^*\text{=}  
 \underset{\{\mathbf{W}_i\}_{i=1}^K }{\text{minimize }}
&   \sum_{i=1}^{K}\text{Tr}[\mathbf{W}_i] \\
  \text{subject to}~\nonumber
&   \text{Tr}[\mathbf{H}_k \mathbf{W}_i]-\gamma_i\sum_{j=1}^{i-1}\text{Tr}[\mathbf{H}_k \mathbf{W}_j] \geq \nonumber \\ & \gamma_i\sigma_k^2,  \forall i\in \mathcal{K}, k\leq i,\label{sdp11}\\
      &    \text{Tr}[\mathbf{H}_i \mathbf{W}_1] \leq\text{Tr}[\mathbf{H}_i \mathbf{W}_2]\leq \cdots \nonumber \\ &\leq\text{Tr}[\mathbf{H}_i \mathbf{W}_K],  \forall i \in \mathcal{K},\\
     &  \mathbf{W}_i=\mathbf{W}_i^H, \mathbf{W}_i\succeq0,  \forall i \in \mathcal{K}.
\end{align}
\end{subequations}

The above problem ${\overset{\sim}{OP_2}}$ is a standard SDP problem \cite{cuma8}, and
therefore, it leads to an optimal solution. However, this solution will
also be the solution to the original P-Min problem in $OP_{2}$,
provided that they are rank-one matrices \cite{cuma10}. The required beamforming
vectors in $OP_{2}$ can be determined by extracting the eigenvector
corresponding to the maximum eigenvalue of this rank-one
matrices. 

As will be discussed in Section~\ref{sec:evaluation}, this SDP problem
always provides rank-one solutions that same to the solutions to
the GEE-Max problem. This confirms the optimality of the solutions
obtained through the proposed SCA algorithm. Note that the P-Min
design for the $OP_2$ cannot be directly employed to solve the
original GEE-Max problem without knowing the achieved SINRs
($\overset{*}\gamma_i $) that maximize the GEE of the system.  On the other hand,  note that the same performance can be achieved in Dinkelbach's algorithm by setting a high accuracy. This can be carried out by setting the termination threshold to zero  in the Dinkelbach's algorithm (i.e., $\varsigma=0 $ in Step 6 of Algorithm 2).

 \subsection{Complexity Analysis of the Proposed Schemes}	
The computational complexities of the proposed algorithms to solve the GEE-Max optimization problem   are defined as follows:
 \subsubsection{The SCA Technique}
  
An iterative algorithm is developed to solve the original GEE-Max optimization problem $OP_1$ by exploiting the SCA technique in which an  approximated  optimization problem provided in (39) is solved in each iteration and the   approximated terms are updated in the next iteration.   In particular, a    standard second-order cone programme (SOCP)    is solved with a number of second-order cone (SOC) and linear constraints. Hence, the worst-case complexity of the SCA technique can be examined through defining the complexity of this
 SOCP, which is solved through the interior-point methods   \cite{point},  \cite{secondd}.  Furthermore, the total number of    constraints associated with this problem     is   $( 2.5K^2+5.5K+6+q_c)$, where $q_c$ is a constant related to the number of constraints that arise due to the relaxation of the exponential constraints in interior-point methods \cite{cvx20}.  Hence, the total number of  iterations that are required to converge to the solution is bounded by  $\mathcal{O}(\sqrt{2.5K^2+5.5K+6+q_c} \log (\frac{1}{\epsilon}))$, where $\epsilon$ is   the required accuracy.  On the other hand, 
 at each iteration, the work required to achieve the solution   is at most $\mathcal{O}(\mathcal{N}^2 \mathcal{M})$ \cite{secondd}, where $\mathcal{N}$ and $\mathcal{M}$ denote the number of optimization variables and the total dimensions of the optimization problems, respectively. For the developed SCA based algorithm, $\mathcal{N}$ and $\mathcal{M}$ are estimated as  
  $( 1.5{K^2} +3.5K +2NK+3+q_c)$ and $( 5.5{K^2} +4K +2NK+4+q_c)$, respectively.

  \subsubsection{The Dinkelbach's Algorithm}
 Now, we define the  computational complexity of the proposed  Dinkelbach's algorithm in which  the  convex parametrized problem provided in (50) is iteratively solved at each iteration for each non-negative parameter $\chi$. In particular, similar to the SCA technique, a standard SOCP with a set of SOC constraints is solved through the interior-point methods. Furthermore, this SOCP mainly determines the computational complexity  of the algorithm.  Note that the number of  constraints in the SCA technique and the Dinkelbach's algorithm are \textit{approximately} the same. Hence,  the  estimated work to determine the solution at each iteration is approximately similar to that required in the SCA based algorithm. However, due to the parametrization required in the Dinkelbach's algorithm, an additional iterative algorithm is required to obtain the optimal $\chi$, as shown in Algorithm 2. The total maximum number of required iterations can be defined by  $\mathcal{O}(\sqrt{4K^2+4K+4+q_c} \log (\frac{1}{\epsilon})\log (\frac{1}{\varsigma}))$, which is higher than that required in the  SCA based algorithm.\\

 \subsection{Convergence Analysis of the Proposed Schemes}
\textbf{SCA Technique:}    The convergence of the 
 SCA-based  technique  proposed to solve the GEE-Max problem  can be examined similar to the extensive anaylsis   presented in \cite{sqa}. In particular, as shown in Algorithm 1, the optimization parameters at the n$^{th}$ iteration (i.e., $\Lambda^{(n)})$ are updated based on the solution that obtained by solving the approximated optimization problem in (\ref{eqn:se12}). In this convergence analysis, three key conditions should be satisfied. First, initializing the optimization problem   with   appropriate initial parameters   $\Lambda^{(0)}$   ensures the feasibility of the problem at each iteration,  which provides a feasible solution to update the parameters for next iteration.  Second, it is obvious that the optimization parameter $\alpha$ is linear, and hence, non-decreasing  (i.e.,  $\alpha^{(n+1)} \geq \alpha^{(n)}$). Finally, as shown in constraint   (9),  the available  power is  upper bounded by $P_{ava}$, such that $P_{ava} << \infty$, which implies that   $\alpha$ is upper bounded, as well. The satisfaction of these three conditions ensures that the developed SCA technique converges to the solution with finite number of iterations. \\


\textbf{The Dinkelbach's Algorithm:}
The convergence analysis of the proposed algorithm based on Dinkelbach's algorithm to solve the GEE-Max problem can be developed similar to that of the SCA technique.   Algorithm 2  consists of two iterative algorithms which are alternately solved to determine the solution to the original GEE-Max problem. In particular, the convergence in term of $\chi$ of the parametrized problem in (50) towards the solution is introduced in Lemma 1. Furthermore, for each $\chi$, the convergence of the developed SCA technique can be proved by following the same procedure as for the convergence of the SCA technique. This confirms the convergence of the proposed algorithm based on Dinkelbach's algorithm. Hence, we can state that the Dinkelbach's algorithm converges to the solution with finite number of iterations.
 
\section{Simulation Results}
\label{sec:evaluation}
In this section, we provide   simulation results to demonstrate the
effectiveness of the proposed GEE-Max algorithms. We   compare the 
performance of these algorithms against the existing
conventional beamforming designs in the literature.

In   simulations, we consider a system with a base station that
communicates with $K$ single-antenna users, which are located at
different distances. The Table~\ref{tab:configa} shows different
parameters we adopted in   simulations.  { Note that these parameters are similar to the parameters used in  \cite{hanif} \cite{ee2}.}

\begin{table}[htb]
\caption{Parameter values used in  the simulations.}
\centering
\label{tab:configa}
\begin{tabular}{|l|r|}
\hline
{\bf Parameter} & {\bf Value(s)}\\
\hline
\hline
Transmit antennas ($N$)  & $3$ \\
User distances  (m)                                 & $[1.0, 5.5, 10.0]$ \\ 
  Path loss exponent ($ \kappa$)  & $1.0$\\
Noise variance of  users ($\sigma^2$) & $2.0$ \\
Threshold for algorithm~\ref{alg:sca}  ($\epsilon$) & $0.01$\\
Threshold for algorithm~\ref{alg:dinkel}  ($\varsigma$) & $0.01$\\
Power-amp efficiencies ($ \epsilon_0)$  &  $0.65$\\
User SINR thresholds for $OP_{1}$ & $10^{-2}$\\
Bandwidth $B_w$  (MHz) & $1$  \\
{Small scale fading $\mathbf{g}_i$  }  &{Rayleigh fading}  \\
\hline
\end{tabular}
\end{table}

\setlength{\textfloatsep}{5pt}

The performance of the system is evaluated in terms of
the achieved EE against different normalized transmit powers. This is
defined by TX-SNR in dB as follows: 

$$\text{TX-SNR
  (dB)}=10\log_{10}\frac{P_{ava}}{\sigma^{2}}. 
$$

\begin{figure}[H]
\centering
\includegraphics[width=.5\textwidth]{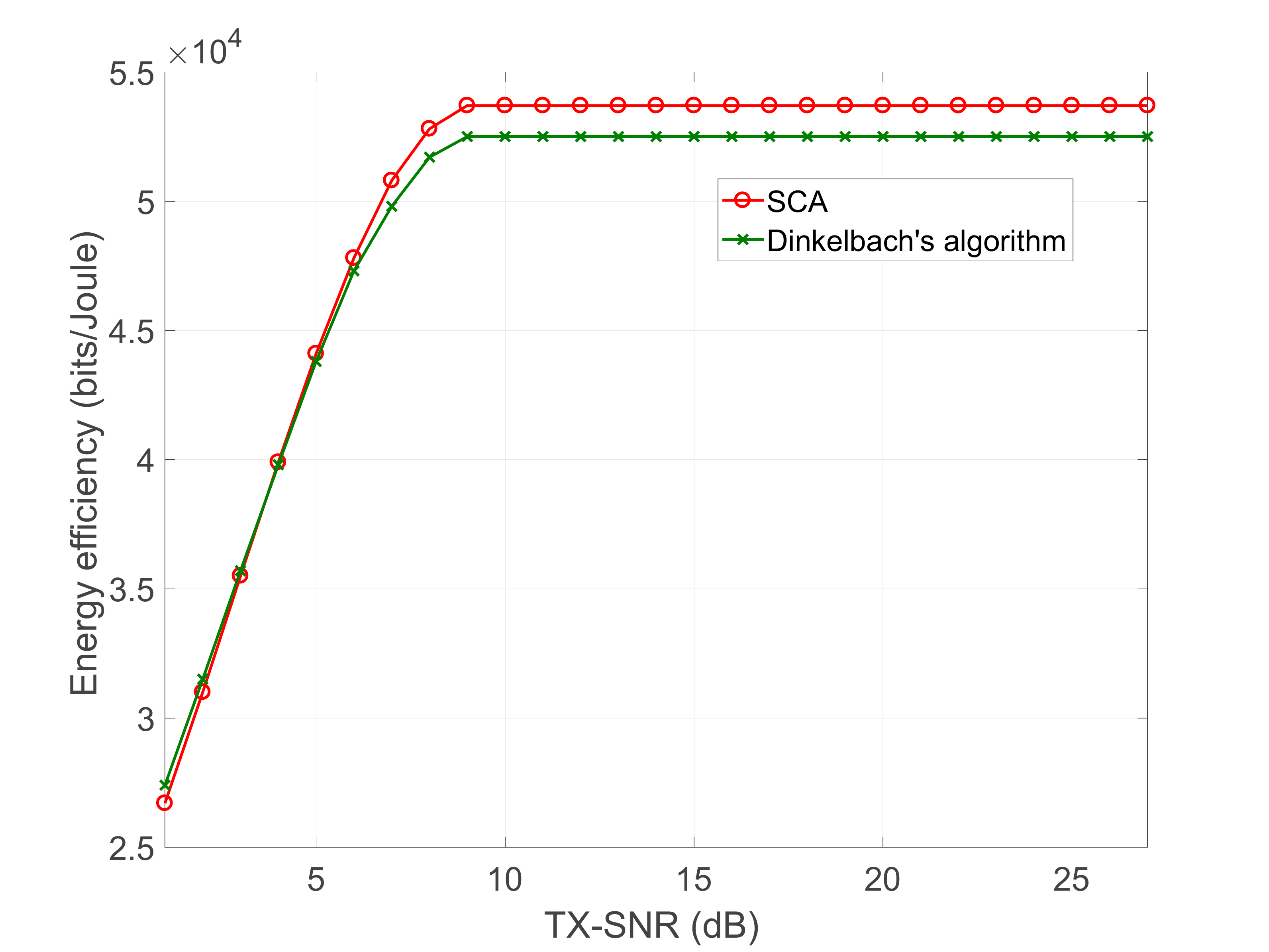}
\caption[width=.5\textwidth]{Achieved EE for GEE-Max-based design through Algorithms~\ref{alg:sca} and~\ref{alg:dinkel}.}
\label{fig:scadin}
\end{figure}

\begin{figure}[t]
 \centering
\includegraphics[width=.5\textwidth]{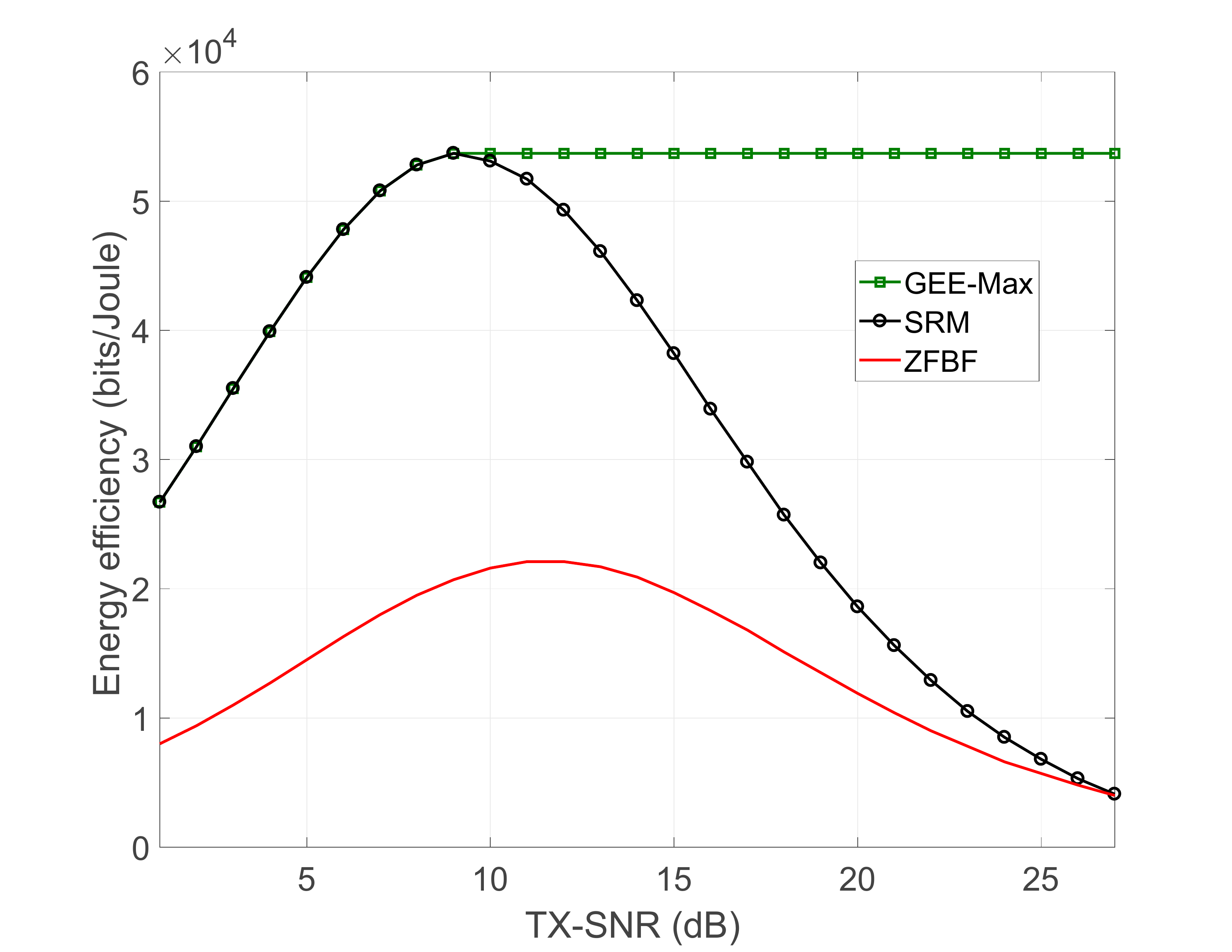}\caption{EE for different design criteria.}
\label{fig:se1ee}
\end{figure}

\begin{figure}[t]
\centering
\includegraphics[width=.5\textwidth]{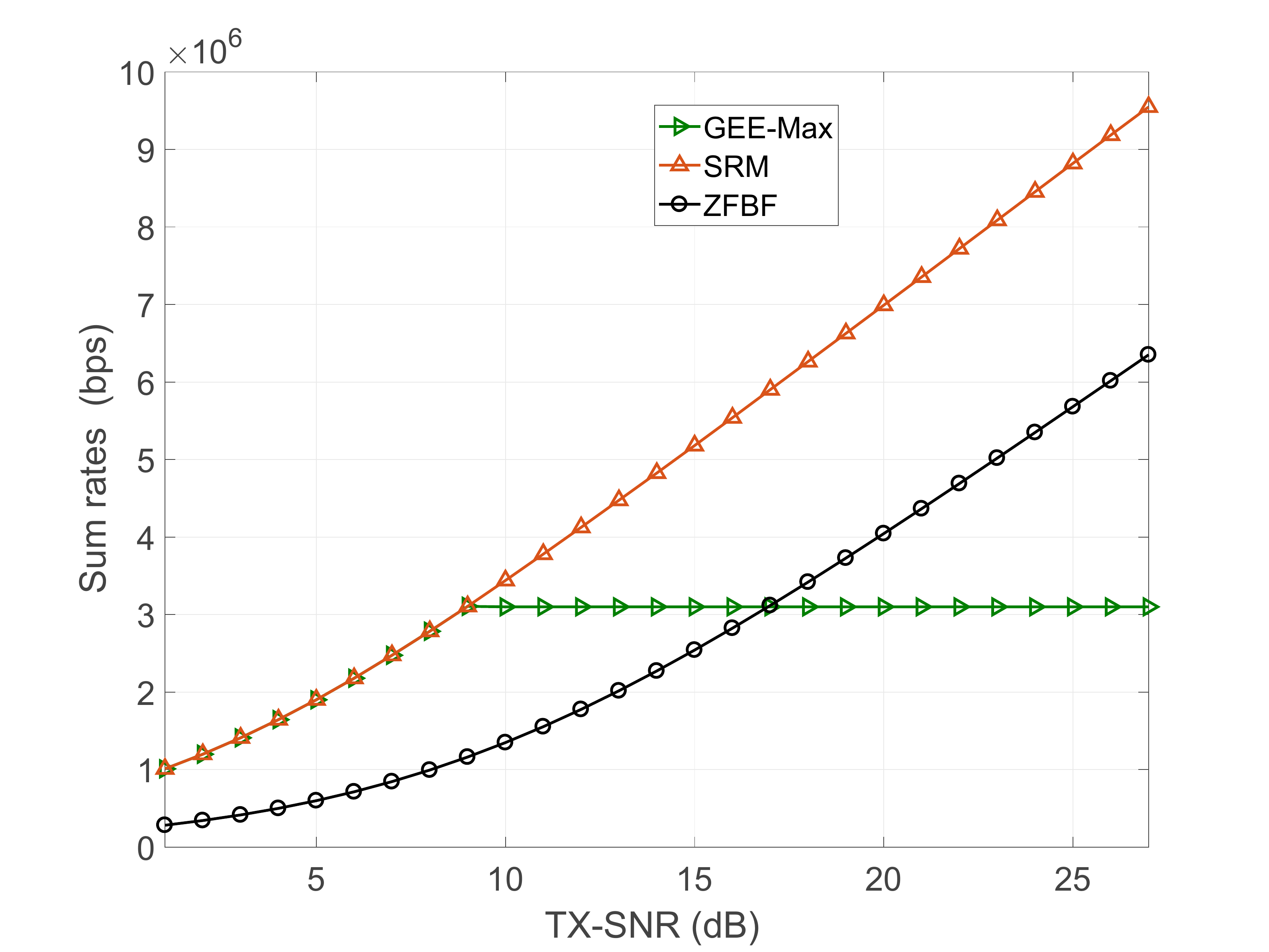}
\caption{Achieved sum rates of different beamforming designs against transmit power.}
\label{fig:rse1ee}
\end{figure}

Figure~\ref{fig:scadin}  shows the achieved  EE for the GEE-Max-based designs with different TX-SNR using the algorithms developed through the SCA and  Dinkelbach's techniques.  The performance gap  between these two approaches is not significant in terms of the achieved EE. However, the design based on the SCA approach outperforms the latter due to the  parametrization of the  objective function in the latter. As seen in  Figure~\ref{fig:scadin}, the achieved EE increases with the available transmit power until it reaches the corresponding maximum green power, where it saturates.\\ 
\indent In order to demonstrate the advantages of the proposed GEE-Max-based design, we compare the EE of the proposed design with the existing conventional  beamforming designs in the literature, namely, beamforming design for SRM in MISO-NOMA system~\cite{hanif},  and maximizing
the sum rate in MISO-OMA~\cite{zf1,zf2} based on  the zero-forcing beamforming designs (ZFBF).  As evidenced by results in Figure~\ref{fig:se1ee}, the GEE-Max based design   outperforms the other designs in terms of achieved EE.  On the other hand, the EE of the  SRM-based designs declines dramatically when the transmit power exceeds the green power. This is particularly the case for the SRM-based designs for both NOMA and OMA (i.e., ZFBF), where these design  consume all  available power for maximizing the achieved sum rate,  as  will be seen below.\\
\indent In order to demonstrate the trade-off between  the achieved EE and the sum-rate   across different beamforming designs, we evaluate the performance of the proposed schemes in terms of the achieved sum-rate of the overall system. Figure~\ref{fig:rse1ee} illustrates the achieved sum-rates of different designs against a range of transmit powers. As expected, the SRM-based design shows the same performance as the GEE-Max design up to the green power, and outperforms the GEE-Max scheme when the  available transmit power exceeds the green power. The sum-rate of the GEE-Max-based scheme remains constant in this region,   where it achieves the maximum EE  as shown in Figure~\ref{fig:se1ee}. On the other hand, the achieved sum rates of both SRM and ZFBF schemes  increases with the available transmit power while decreasing their of  EE performance~(Figure~\ref{fig:se1ee}).

\begin{table*}
\centering
\caption { Power allocations and the achieved SINRs for different GEE-Max designs using the  SCA, TX-SNR= 2dB.} 
\label{tab:powersinr1}
 \centering
 \begin{tabular}{|c|c|c|c|c|c|c|c|c|c|}
    \hline
    \multirow{2}{*}{Channels} &
      \multicolumn{2}{c|}{User 1}  &
      \multicolumn{2}{c|}{User 2}  &
      \multicolumn{2}{c|}{User 3}\\
      \cline{2-7}
    &$\overset{*}\gamma_1$ &  $P_1$ (W) &$\overset{*}\gamma_2$ & $P_2$ (W)& $\overset{*}\gamma_3$ &  $P_3$ (W)\\
    \hline
     Channel 1 &  1.2468   & 0.9095 & 0.1848 & 0.9095  &0.1482  & 1.3507   \\
    \hline
    Channel  2 & 0.9975 &0.8660 & 0.1535 & 0.8660 &    0.1381  & 1.4378 \\
    \hline
    Channel   3 & 1.3353 &0.9115 &  0.1789 &0.9115 & 0.1512 & 1.3082  \\
    \hline
    Channel 4 & 1.8190 &0.9815  & 0.2485 &0.9815 & 0.1640 & 1.2068  \\
    \hline
     Channel 5 & 1.4606 &0.9400  & 0.2098  & 0.9400  & 0.1551  & 1.2898  \\
    \hline
  \end{tabular}
\end{table*}

\begin{figure}[t]
\centering
\includegraphics[width=.5\textwidth]{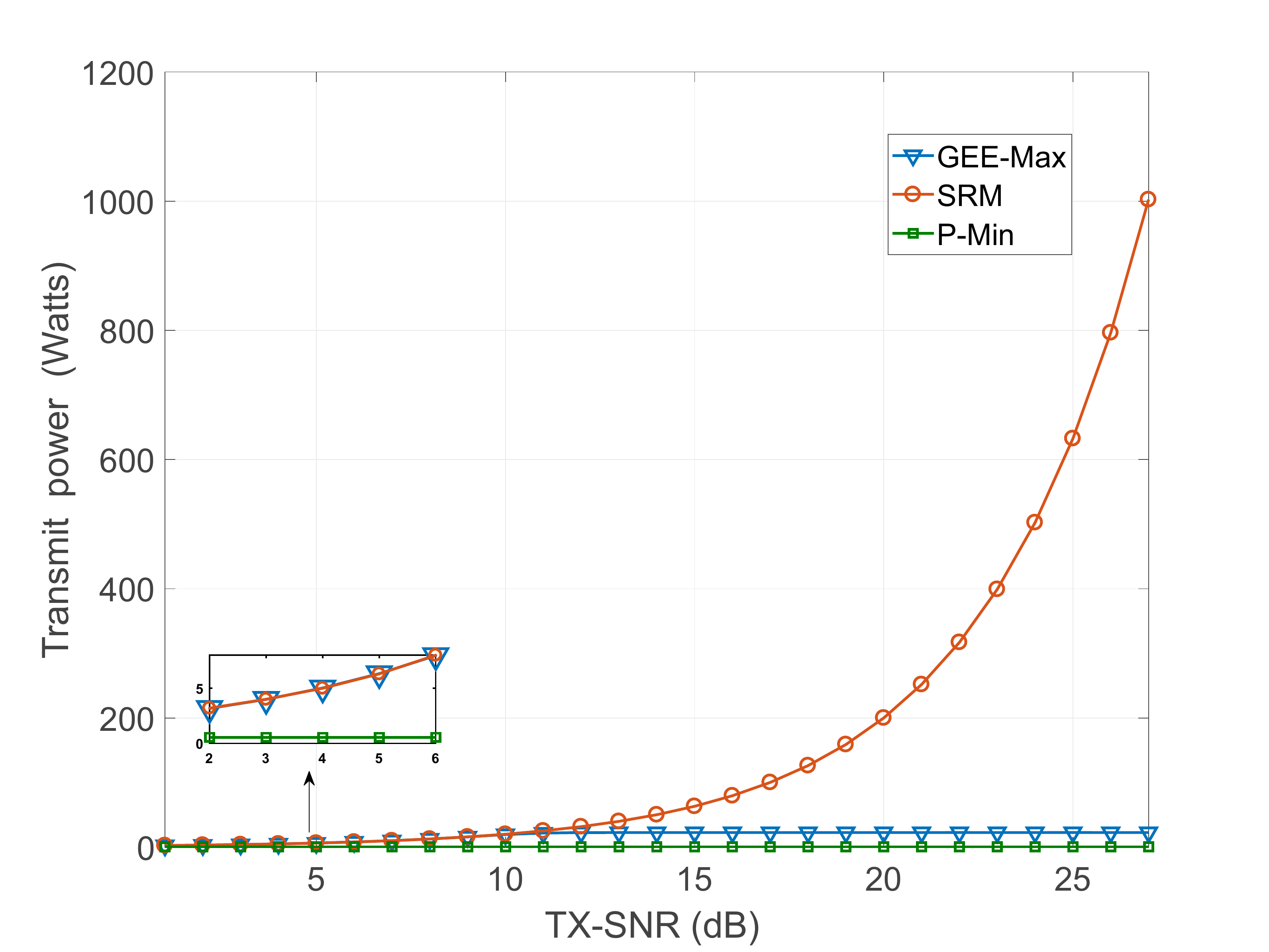}
\caption{Required transmit power for different beamforming design criteria.  }
\label{fig:ploess}
\end{figure}

To further evaluate the transmit power consumption (i.e., $ P_{tr}$),  we compare the transmit power requirements for  different NOMA beamforming designs. As shown in Figure~\ref{fig:ploess}, the  P-Min  beamforming design~\cite{misonoma} outperforms the  SRM- and GEE-Max-based designs. This is because the P-Min-based beamforming design uses the transmit power to satisfy the required SINR constraints. On the other hand, the  SRM-based scheme makes use of all the available transmit power to achieve the maximum sum rate, while the  GEE-Max-based scheme consumes a certain amount of transmit power (i.e., green power) to maximize the GEE of the system. From these observations, the GEE-Max-based design can be considered as the scheme that strikes a good balance between the SRM  and P-Min-based designs.

\begin{figure}[t]
 \centering
\includegraphics[width=.5\textwidth]{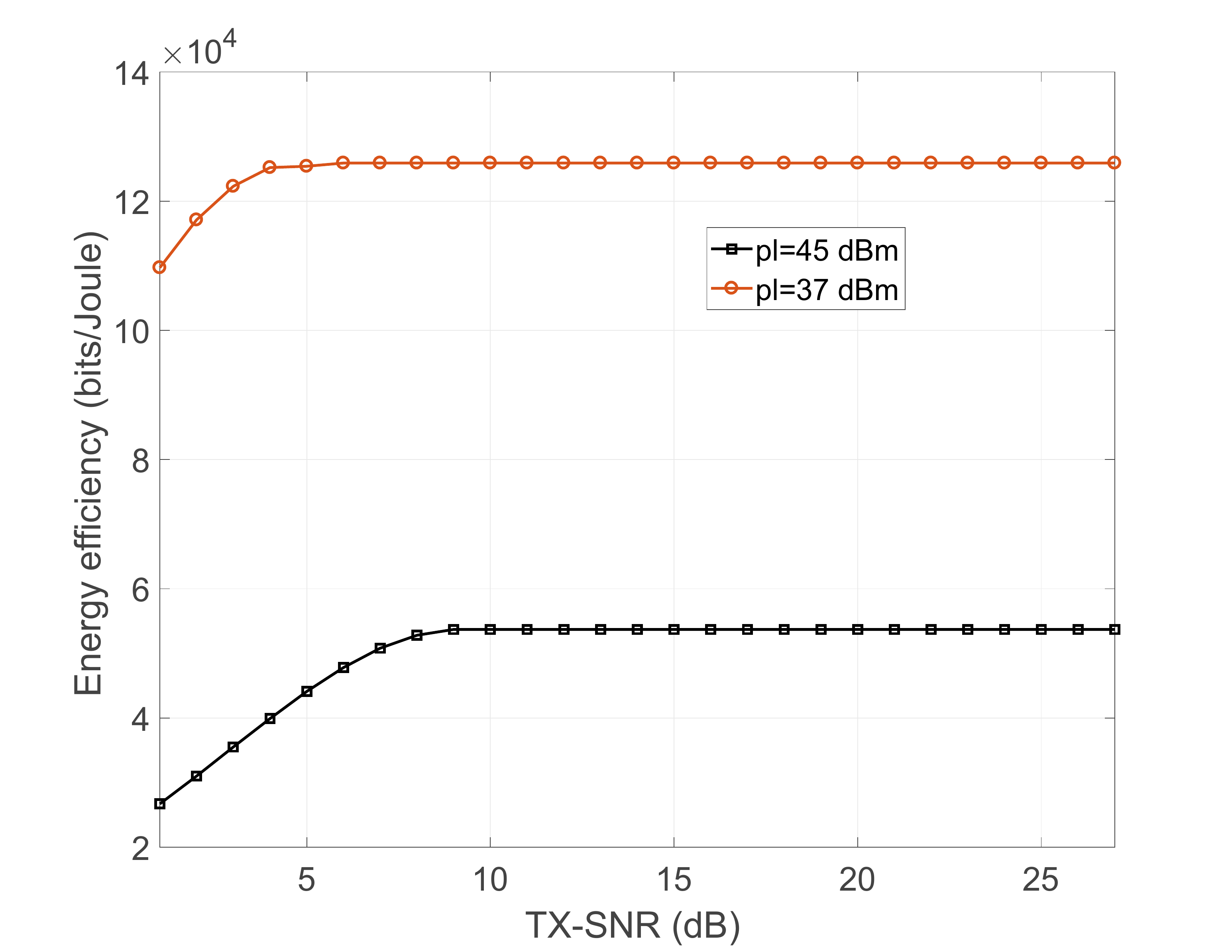}
\caption{Achieved EE of GEE-Max design with different power losses.}
\label{fig:ploess1}
\end{figure}

Furthermore, we evaluate the impact of the power losses on the performance of the proposed GEE-Max  design. Figure~\ref{fig:ploess1} shows the achieved EE against different power losses. There are two key observations to be drawn from  Figure~\ref{fig:ploess1}. Firstly, the achievable EE decreases as the power loss increases. Secondly, the green power that achieves the maximum EE increases as the power losses increases.

\begin{table}[htb]
\centering
\caption { Power allocations for the achieved SINRs in Table~\ref{tab:powersinr1} using the P-Min  design.} \label{tab:powersinr2}
  \begin{tabular}{|l|l|l|l|l|l|l|l|l|l|}
  \hline
    \multirow{2}{*}{Channels} &
      \multicolumn{3}{c|}{$P_i$=Tr[$\mathbf{W}_i]$}\\
      \cline{2-4}
    & $P_1$ (W)&$P_2$ (W)&$P_3$ (W)\\
    \hline
   Channel  1 &  0.9094&0.9095& 1.3508  \\
    \hline
    Channel  2 &  0.8660&   0.8660& 1.4379 \\
    \hline
    Channel  3 & 0.9113 &0.9115& 1.3082  \\
    \hline
   Channel  4 & 0.9815&0.9815& 1.2068 \\
    \hline
    Channel  5 &  0.9400   & 0.9400 & 1.2897 \\
    \hline
  \end{tabular}
\end{table}

Next, we provide   results to validate the optimality of the  proposed SCA-based GEE-Max algorithm. We compare the achieved SINRs and power allocations of the proposed scheme with the P-Min-based scheme, which assumes to use the same SINR targets   obtained in the GEE-Max-based scheme. The power allocations and the achieved SINRs using the proposed SCA- and GEE-Max-based schemes are given for five different random channels in Table~\ref{tab:powersinr1}.   For the same set of channels used in Table~\ref{tab:powersinr1},  the power allocations obtained through solving  the P-Min   ${\overset{\sim}{OP_2}}$  are given in Table \ref{tab:powersinr2} where the achieved SINRs in Table~\ref{tab:powersinr1} have been set as the target SINRs in ${\overset{\sim}{OP_2}}$.   By  comparing the results provided in
Tables~\ref{tab:powersinr1} and~\ref{tab:powersinr2}, it can be concluded that both problems provide the same solutions in terms of power allocation. It can also be  noticed that the beamforming vectors obtained in both case are the same, and they are not presented here for the reasons of brevity. Therefore, we can confirm that the SCA algorithm yields the optimal solution to the original GEE-Max problem within a few cycles of iterations.

{ Now, we consider the effect of the path loss exponent $\kappa$ on the achieved EE for the GEE-Max design in Table \ref{tab:new}. As expected, the achieved EE decreases as $ \kappa$ increases. On the other hand, the base station requires additional transmit  power (i.e., $P_{tr}$) to maximize EE when the path loss exponent $\kappa$ increases, as shown in Table \ref{tab:new}.}

\begin{table}[H]
\centering
\caption { {Achieved  EE  of the proposed GEE-Max design with different path loss exponents $ \kappa$ under TX-SNR= 25 dB.}} \label{tab:new}
  \begin{tabular}{|c|c|c|c|c|c|c|c|c|}
  \hline
    \multirow{2}{*}{Path Loss Exponent  ($ \kappa$)} &
      \multicolumn{2}{c|}{ GEE-Max design}\\
      \cline{2-3}\centering
    & Achieved  EE (Mbits/Joule)&$P_{tr}$ (W) \\
    \hline
      1 &   0.0532  & 22.6202  \\
    \hline
      2 & 0.0450 &  24.9749    \\
    \hline
      3 &0.0254		  &64.2730    \\
    \hline
   4 &0.0101  & 287.4765  \\
    \hline
  \end{tabular}
\end{table}

\indent{ In addition, Figure~\ref{fig:antt}  illustrates the achieved EE against different number of transmit antennas for the proposed GEE-Max and the SRM designs. In general, the increase in the number  of  the transmit antennas  provides additional degrees of freedom, and hence, 
improves the achieved sum-rate of the system through efficient interference mitigation \cite{ee2}. However, the achieved EE shows a different performance behaviour,  as the increase of $N$ will also increase $P_{loss}$, accordingly.  With the GEE-Max design,  two  different behaviours can be observed in Figure~\ref{fig:antt}, as follows. First, the achievable EE increases with the number of  transmit antennas. This is due to the fact that the rate improvement offered by the additional number of antennas is dominant in achieved EE than the power loss (i.e., the increase of $P_{loss}$ due to the increase of $N$) introduced by those antennas.  Hence, the achieved EE increases gradually until it reaches its  maximum with  $N=3$. Then, the power loss with more number of transmit antennas becomes more dominant in the achieved EE than the rate improvement offered by those antennas. Hence, with more number of antennas, the achieved EE begins to decrease and shows the same performance as in the SRM design as seen in Figure~\ref{fig:antt}.  However, the EE achieved through the GEE-Max design outperforms   the SRM design.  Based on these observations, we can conclude that  there is an optimal number of transmit antennas which can achieve  the maximum EE  and employing a larger number of antennas than the optimal number   will introduce a loss in the achievable EE performance of the system.   Note that this optimal number of transmit antennas depends on the system parameters (i.e., $K$, $p_{dyn}$, $p_{sta}$, etc.). Furthermore, both the SRM design and GEE-Max design achieve the same EE when the number of  transmit antennas is greater than 7. This is due to the fact that the available power (i.e., $P_{ava}$) is less than the green power, hence, both designs will provide the same solution and achieve the same EE performance.  

\begin{figure}[H]
\centering
\includegraphics[width=.5\textwidth]{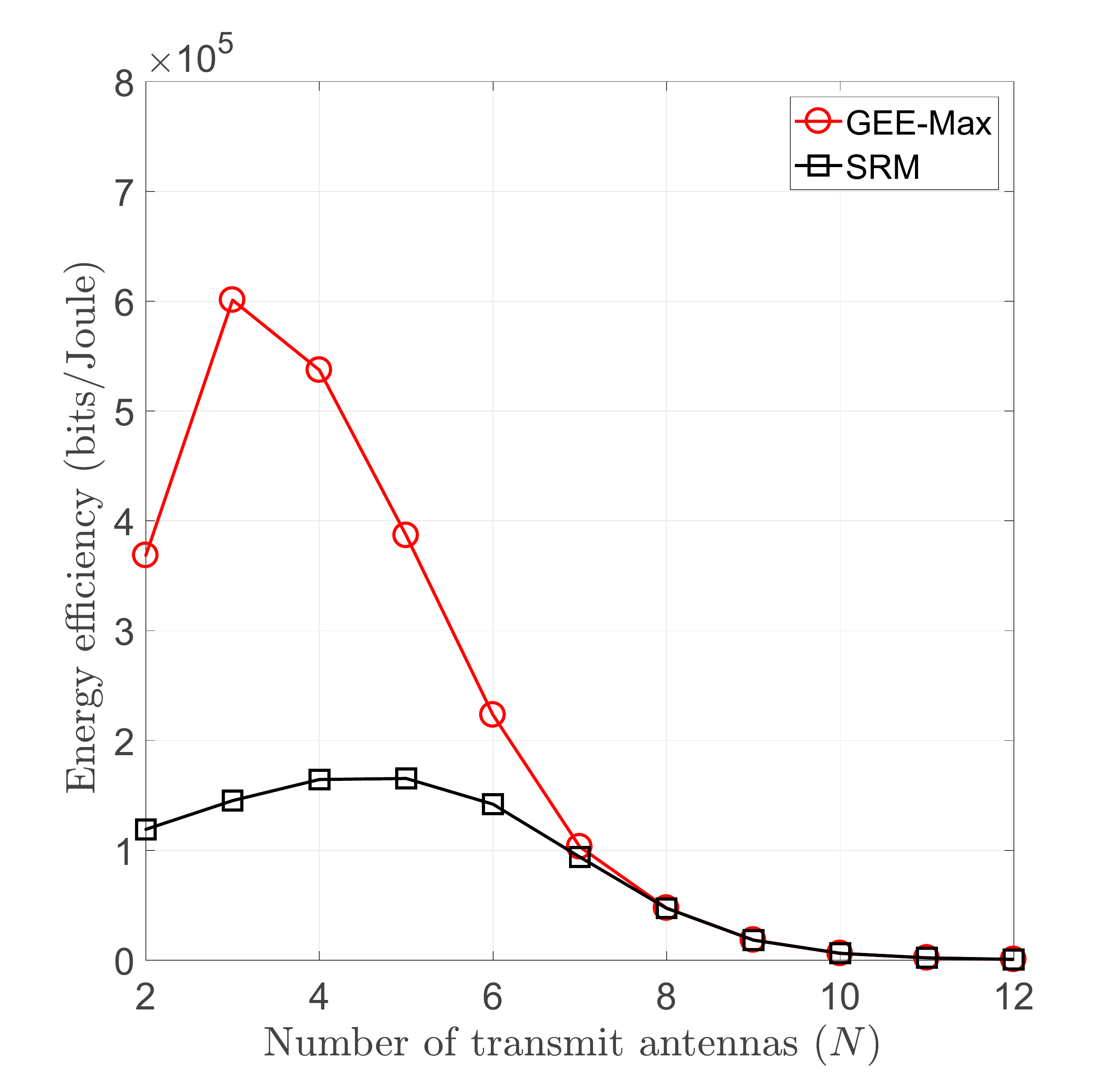}
\caption{The achieved EE against different number of transmit antenna $N$ for GEE-Max and SRM designs at TX-SNR =10 dB. The $p_{sta}$ and $p_{dyn}$ are set to be 10 dBm and 5 dBm, respectively. }
\label{fig:antt}
\end{figure}

 }

\begin{figure}[H]
\centering
\includegraphics[width=.5\textwidth]{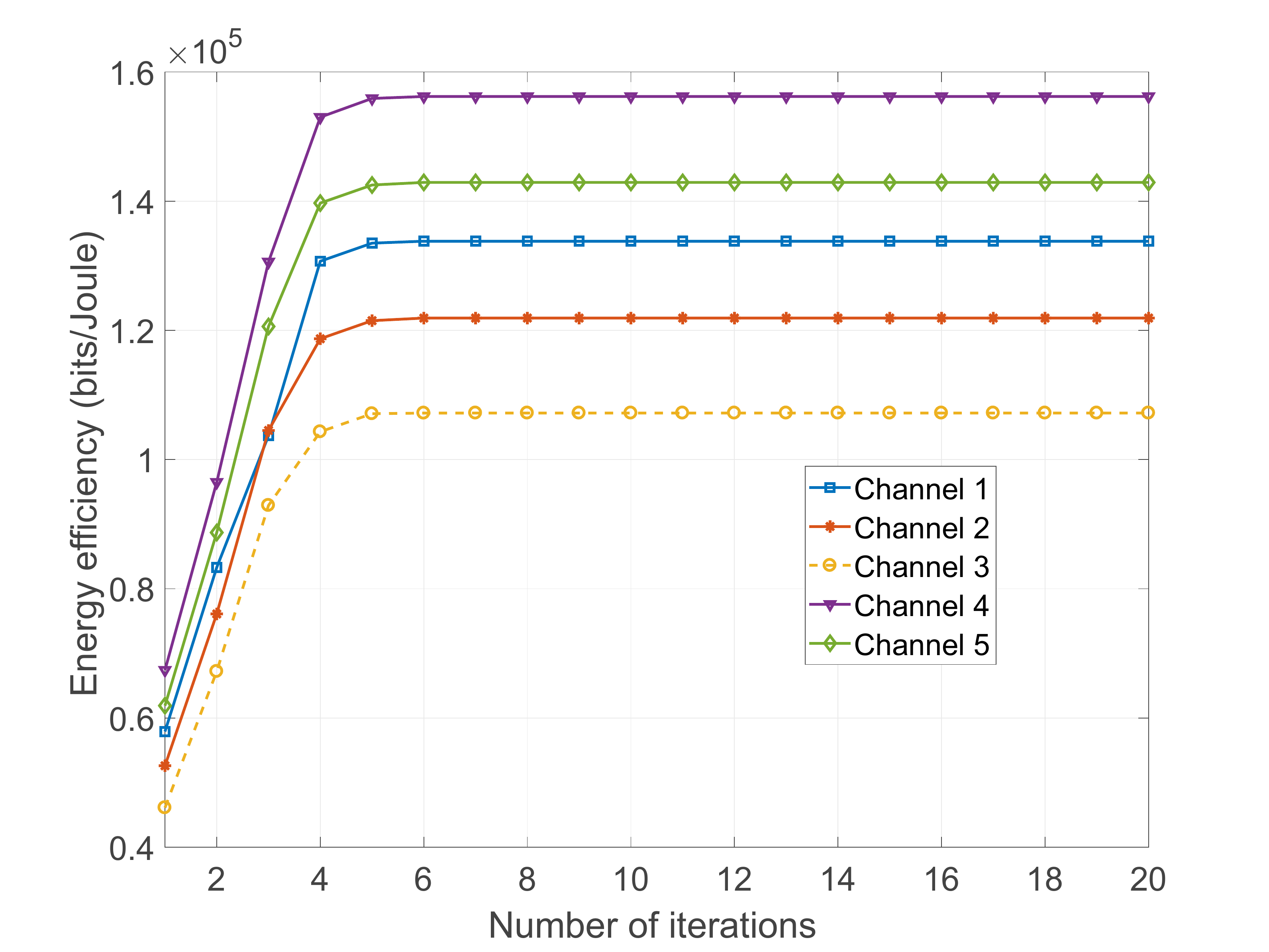}
\caption{The convergence of SCA based GEE-Max algorithm for different set of random channels. The TX-SNR and $P_{loss}$ are set to be 20 dB and 40 dBm, respectively. }
\label{fig:conv}
\end{figure}

Finally, we  evaluate  the number of iterations required for the  convergence of the proposed SCA-based GEE-Max   algorithm. Figure~\ref{fig:conv}  depicts the convergence of the proposed SCA algorithm with different set of channels. The threshold ($\varepsilon $ in Algorithm~\ref{alg:sca}) to terminate the algorithm has been set to 0.01. As seen in Figure~\ref{fig:conv}, the algorithm  converges within a few  iterations.

\section{Conclusions}
\label{sec:conclusions}

In this paper, we proposed two different algorithms for   energy efficient beamforming designs to maximize the GEE  for  MISO-NOMA  systems. These two algorithms stem from an approach whereby we transformed the original, non-convex GEE-Max optimization problem into an approximated convex problem. The  algorithms are based on the sequential convex programming and the  Dinkelbach's approaches, respectively.  Our evaluation, benchmarked against a baseline, showed that the proposed algorithms   converge and produce similar results, and    also outperform the benchmark approaches. Our evaluation also verified the   optimality of the proposed SCA-based design  by comparing  the  power allocations with the an equivalent P-Min design with the same set of random channels. Furthermore, it was shown that the GEE-Max based design is capable of achieving a good trade-off between the designs with conflicting performance metrics that maximize  the sum rate and minimize  the transmit power.

\section*{Appendix A}

\section*{PROOF OF THEOREM 1}
%
  To prove  Theorem 1, we firstly rewrite (\ref{eqn:Nnn}) such as
\begin{equation}\label{popo}
\chi^*=  \frac{f_1(\{\mathbf{w}^*_i \}_{i=1}^{K} )}{f_2(\{\mathbf{w}^*_i \}_{i=1}^{K} )} \geq \frac{f_1(\{\mathbf{w}  _i \}_{i=1}^{K} )}{f_2(\{\mathbf{w}  _i \}_{i=1}^{K} )},
\end{equation}
 where $\{\mathbf{w}_i^*  \}_{i=1}^{K}$ denote  the beamforming vectors that maximize the original problem $OP_1$. Without loss of generality, the condition in (\ref{popo}) can be decomposed as
\begin{subequations}
\begin{align}
f_1(\{\mathbf{w} _i \}_{i=1}^{K} )-\chi^*f_2(\{\mathbf{w} _i \}_{i=1}^{K} )\leq0,\label{koop}\\
f_1(\{\mathbf{w}^*_i \}_{i=1}^{K} )-\chi^*f_2(\{\mathbf{w}^*_i \}_{i=1}^{K} )=0, \label{koop1}
\end{align}
\end{subequations}
where the left side of  (\ref{koop}) denotes the objective function of the  parametrized optimization problem  $OP_5$ (i.e., $F(\{\mathbf{w}_i \}_{i=1}^{K},\chi^*)$).
 The inequality in (\ref{koop})    reveals  that any feasible beamforming set $\{\mathbf{w}_i\}_{i=1}^{K} $ (rather than the optimal set) will provide $F(\{\mathbf{w}_i \}_{i=1}^{K},\chi^*)$ to be less than zero, whereas  the optimal beamfroming vectors $ \{\mathbf{w}_i^* \}_{i=1}^{K}$ could be achieved if  and only if the condition in (\ref{koop1}) is satisfied. Hence, we can determine the optimal beamforming vectors of the original fractional problem $OP_1$ by solving the non-fractional one  in $OP_5$ with the assumption that the maximum objective value  of $OP_5$ is zero.    This completes the proof of Theorem 1.  \hfill$\blacksquare$

 \section*{Appendix B}
\section*{PROOF OF LEMMA 1 }
 In order to prove  the   convergence of the Dinkelbach's iterative approach to  the optimal solution, the following conditions can be equivalently proven \cite{dink}:
 \begin{subequations} \label{pop}
 \begin{align}
    \chi^{(n+1)} \geq \chi^{(n )} ,\label{sdrt}\\
  \underset{n\rightarrow\infty}{\text{lim}}\chi^{(n )}=\chi^* \label{kawww}.
   \end{align}
\end{subequations}
 We start with $ \chi^{(n+1)} \geq \chi^{(n )}$, and it is known that    $F(\chi)$ is a  non-decreasing function. Therefore
 $$ \{F(\chi^{(n)})\geq F(\chi^{*}) \geq 0 |  \chi^{(n )} \leq  \chi^{*}\}, $$
 which implies that
 \begin{equation}\label{kha}
 f_1(\{\mathbf{w}^{(n )}_i \}_{i=1}^{K} )-\chi^n f_2(\{\mathbf{w}^{(n )}_i \}_{i=1}^{K} ) \geq 0.
\end{equation}
On the other hand,    the following holds based on (\ref{eqn:taha}):
\begin{equation}\label{kao}
f_1(\{\mathbf{w}^n_i \}_{i=1}^{K} )=\chi^{(n+1)} f_2(\{\mathbf{w}^{(n )}_i \}_{i=1}^{K} ).
\end{equation}
 By substituting (\ref{kao})  in (\ref{kha}),   we have:
 $$ (\chi^{(n+1)}- \chi^{(n )}) f_2(\{\mathbf{w}^{(n )}_i \}_{i=1}^{K} ) > 0.$$
Since  $f_2(\{\mathbf{w}^{(n )}_i \}_{i=1}^{K} )$ is   assumed to be always positive, then
 $$ (\chi^{(n+1)}- \chi^{(n )})  > 0,$$
 which confirms the inequality in  (\ref{sdrt}).
Now, we consider the second condition in (\ref{kawww}) and  prove this through contradiction. First, we assume that the condition in  (\ref{kawww}) does not hold and there exists another non-negative parameter  ($ \chi^+$) such that \\
  $$\underset{n\rightarrow\infty}{\text{lim}}\chi^{(n )}=\chi^+ <\chi^*.$$
 Based on this argument, the following holds:
 $$ F(\chi^+)=0.$$
However,  $F(\chi )$ is a non-decreasing function, which means that
\begin{equation}
 \{F(\chi^+)=0>F(\chi^*)=0 | \chi^+<\chi^*\},
\end{equation}
   which is obviously  not true and contradicts   the assumption   made  at the beginning of this proof.  Therefore,
   $$\underset{n\rightarrow\infty}{\text{lim}}\chi^{(n )}=\chi^*.$$
This confirms that the Dinkelbach's iterative algorithm converges to the optimal solution, which completes the proof of Lemma 1.\hfill$\blacksquare$

\bibliographystyle{IEEEtran}
\bibliography{references}
\begin{IEEEbiography}[{\includegraphics[width=1in,height
=1.25in,clip,keepaspectratio]{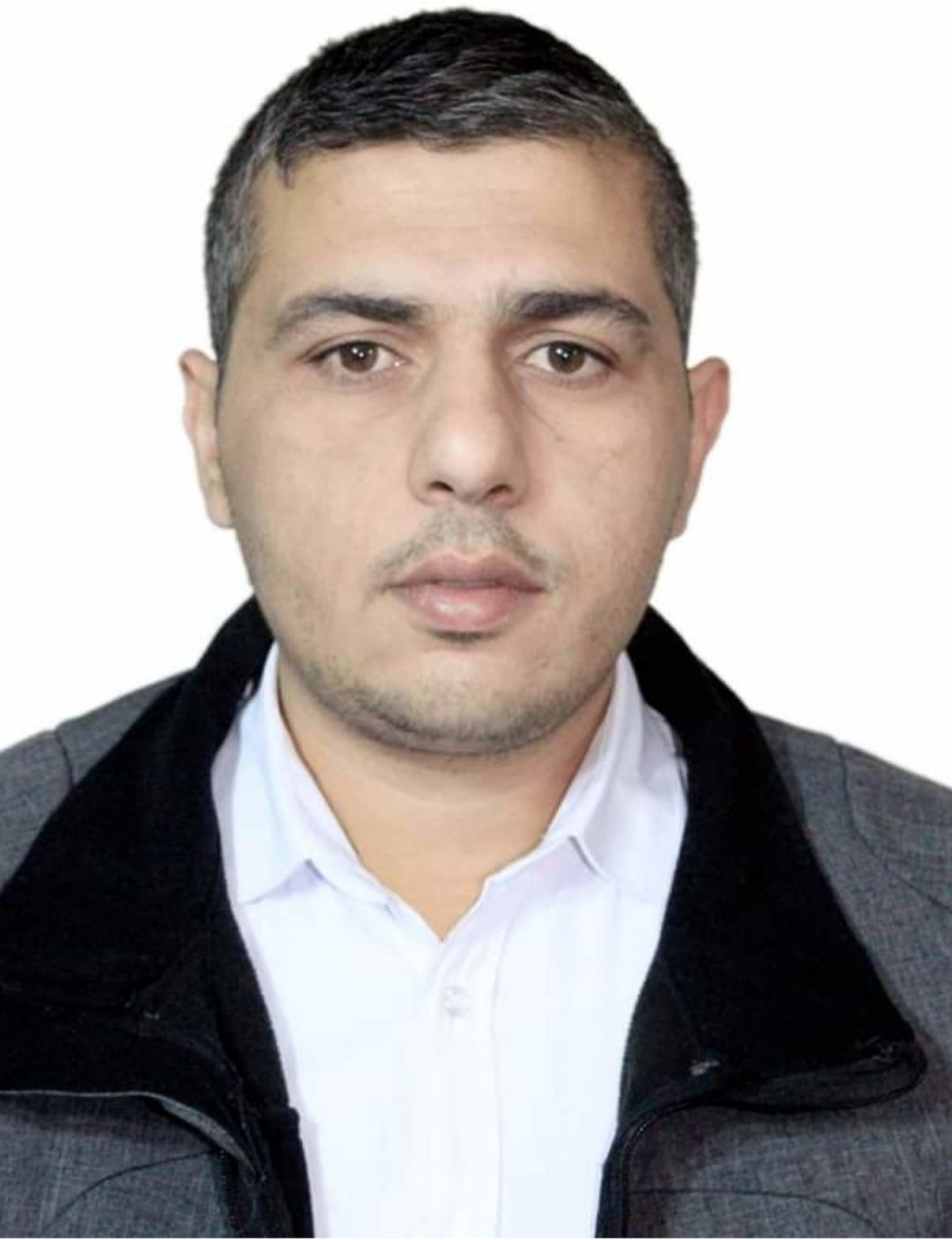}}]{Haitham Moffaqq Al-Obiedollah} received the
BSc    degree  from the Electrical Engineering Department, Jordan University of Science
and Technology, Jordan in 2006  and Ms.c in communication engineering from Yarmouk University, Jordan in 2012.  He is currently pursuing
the Ph.D. degree with the Department of Electronics  Engineering, the 	 University of York, U.K.
He is funded by Hashemite University, Jordan. His
current research interests include non-orthogonal multiple access (NOMA), resource allocation techniques, beamforming designs, multi-objective optimization techniques, and convex optimization theory.
\end{IEEEbiography}

 \begin{IEEEbiography}[{\includegraphics[width=1in,height
=1.25in,clip,keepaspectratio]{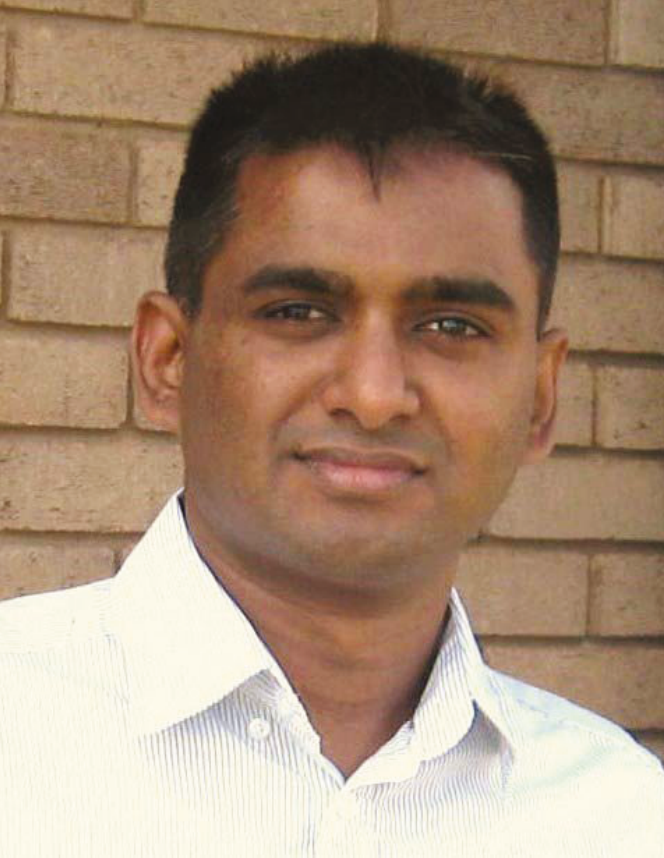}}]{Kanapathippillai Cumanan}(M'10)  received the B.Sc. (Hons.) degree in electrical and electronic engineering from the University of Peradeniya, Sri Lanka, in 2006, and the Ph.D. degree in signal processing for wireless communications from Loughborough University, Loughborough, U.K.,in 2009.

He is currently a Lecturer with the Department of Electronics, University of York, U.K. He was with the School of Electronic, Electrical and System Engineering, Loughborough University, U.K. He was a Teaching Assistant with the Department of Electrical and Electronic Engineering, University of Peradeniya, Sri Lanka, in 2006. In 2011, he was an Academic Visitor with the Department of Electrical and Computer Engineering, National University of Singapore, Singapore. He was a Research Associate with the School of Electrical and Electronic Engineering, Newcastle University, U.K., from 2012 to 2014. His research interests include physical layer security, cognitive radio networks, relay networks, convex optimization techniques, and resource allocation techniques.

Dr. Cumanan was a recipient of an Overseas Research Student Award Scheme from Cardiff University, Wales, U.K., where he was a Research Student from 2006 to 2007.
\end{IEEEbiography}

\begin{IEEEbiography} 
{ Jeyarajan Thiyagalingam} received his PhD degree in Computer Science from Imperial College, London, in 2005. Currently, he is a Senior Scientist at the Science and Technologies Facilities Council, Rutherford Appleton Laboratory (STFC-RAL) in Harwell, Oxford, UK. Before joining STFC-RAL, he was an academic at the University of Liverpool, and previously held positions at MathWorks UK and at the University of Oxford. His research interests include  computationally efficient algorithms and models for machine learning systems, machine learning models for signal processing applications, such as target tracking, estimation, and application of machine learning for fundamental science.  He is a Fellow of the British Computer Society and also a member of IET and IEEE.
\end{IEEEbiography}

 \begin{IEEEbiography}[{\includegraphics[width=1in,height
=1.25in,clip,keepaspectratio]{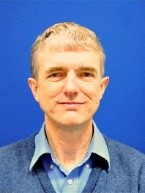}}] {Alister Burr} was born in London, U.K, in 1957.  He received the BSc degree in Electronic Engineering from the University of Southampton, U.K in 1979 and the PhD from the University of Bristol in 1984.  Between 1975 and 1985 he worked at Thorn-EMI Central Research Laboratories in London.  In 1985 he joined the Department of Electronics (now Electronic Engineering) at the University of York, U.K, where he has been Professor of Communications since 2000.  His research interests are in wireless communication systems, especially MIMO, cooperative systems, physical layer network coding, and iterative detection and decoding techniques.  He has published around 250 papers in refereed international conferences and journals, and is the author of “Modulation and Coding for Wireless Communications” (published by Prentice-Hall/PHEI), and co-author of “Wireless Physical-Layer Network Coding (Cambridge University Press, 2018).  In 1999 he was awarded a Senior Research Fellowship by the U.K. Royal Society, and in 2002 he received the J. Langham Thompson Premium from the Institution of Electrical Engineers.  He has also given more than 15 invited presentations, including three keynote presentations.  He was chair, working group 2, of a series of European COST programmes including IC1004 “Cooperative Radio Communications for Green Smart Environments” (which have been influential in 3GPP standardisation), and has also served as Associate Editor for IEEE Communications Letters, Workshops Chair for IEEE ICC 2016, and TPC co-chair for PIMRC 2018.
\end{IEEEbiography}

\begin{IEEEbiography}[{\includegraphics[width=1in,  height=1.25in,clip, keepaspectratio]{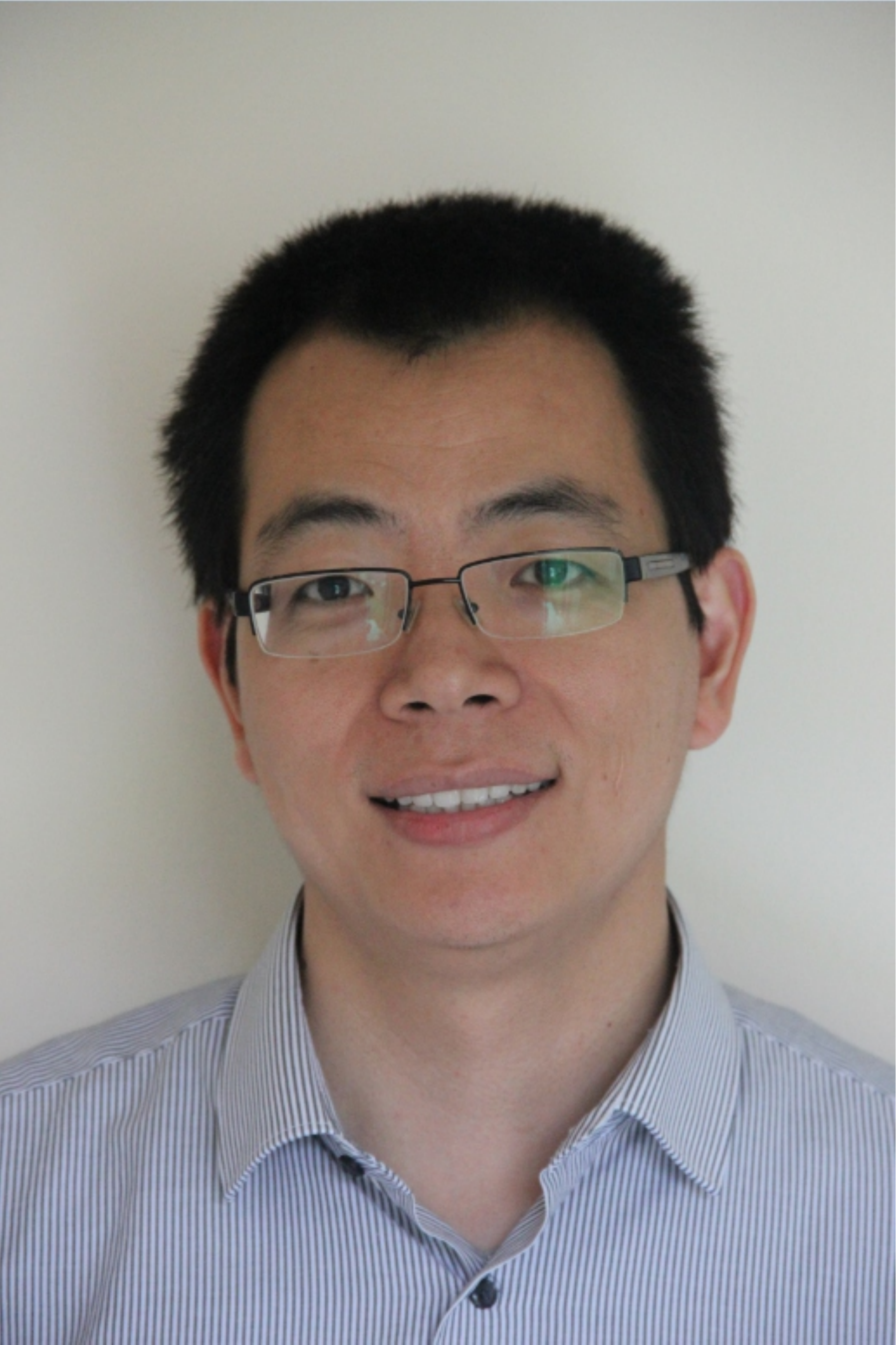}}]{Zhiguo Ding} (S'03-M'05) received his B.Eng in Electrical Engineering from the Beijing University of Posts and Telecommunications in  2000, and the Ph.D degree in Electrical Engineering from Imperial   College London in 2005.   From Jul. 2005 to  Apr. 2018, he was working in Queen's University Belfast, Imperial   College, Newcastle University and Lancaster University. Since Apr. 2018, he has been  with the University of Manchester as a  Professor in Communications. From Oct. 2012 to Sept. 2018, he has also been an academic visitor in Princeton University.  Dr Ding' research interests are 5G networks, game theory, cooperative and energy harvesting networks and statistical signal processing.  He is serving  as an Editor   for {\it IEEE Transactions on Communications}, {\it IEEE Transactions on Vehicular Technology},  and {\it Journal of Wireless Communications and Mobile Computing}, and was an Editor for {\it IEEE Wireless Communication Letters}, {\it IEEE Communication Letters} from 2013 to 2016.  He received the best paper   award in IET ICWMC-2009 and IEEE WCSP-2014,   the EU Marie Curie Fellowship 2012-2014, the Top IEEE TVT Editor 2017, IEEE Heinrich Hertz Award 2018, the IEEE Jack Neubauer Memorial Award 2018 and the IEEE Best Signal Processing Letter Award 2018. 
 \end{IEEEbiography} 

\begin{IEEEbiography}[{\includegraphics[width=1in,  height=1.25in,clip, keepaspectratio]{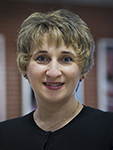}}]
{Octavia A. Dobre} (M’05–SM’07) received the Dipl. Ing. and Ph.D. degrees from Politehnica
University of Bucharest (formerly Polytechnic Institute of Bucharest), Romania, in 1991 and 2000,
respectively. Between 2002 and 2005, she was with Politehnica University of Bucharest and New
Jersey Institute of Technology, USA. In 2005, she joined Memorial University, Canada, where she is
currently Professor and Research Chair. She was a Visiting Professor with Université de Bretagne
Occidentale, France, and Massachusetts Institute of Technology, USA, in 2013.
Her research interests include 5G enabling technologies, blind signal identification and parameter
estimation techniques, as well as optical and underwater communications. She co-authored more than
250 papers in these areas.\\
Dr. Dobre serves as the Editor-in-Chief of the IEEE COMMUNICATIONS LETTERS, as well as an
Editor of the IEEE COMMUNICATIONS SURVEYS AND TUTORIALS and IEEE SYSTEMS. She
was an Editor, Senior Editor, and Guest Editor for other prestigious journals. She served as General
Chair, Tutorial Co-Chair, and Technical Co-Chair at numerous conferences. Dr. Dobre was a Royal
Society Scholar in 2000 and a Fulbright Scholar in 2001. She is a Fellow of the Engineering Institute
of Canada.
 \end{IEEEbiography}

\end{document}